\documentclass[sigconf]{acmart}

\usepackage{microtype}
\usepackage{graphicx}
\usepackage{subfig}
\usepackage{booktabs} 
\usepackage{xtab}
\usepackage{multirow,multicol}
\usepackage{threeparttable}
\usepackage{tablefootnote}
\usepackage{makecell}

\usepackage{amsmath}
\usepackage{amsfonts}
\usepackage{amssymb}
\usepackage{amsthm}
\usepackage{mathtools}

\usepackage{soul}
\usepackage{enumitem}
\usepackage{pifont}
\usepackage{comment}
\usepackage{array}
\usepackage{colortbl}
\usepackage{url}
\usepackage{color}
\usepackage{graphics}
\usepackage{wrapfig}
\usepackage{algorithmic}

\usepackage{algorithm}
\usepackage{float}
\usepackage{tablefootnote}
\usepackage{multirow}

\usepackage{pifont}
\newcommand\mynuma[1]{\ifcase#1 \or \ding{172}\or \ding{173}\or
  \ding{174}\or \ding{175}\or \ding{176}\or \ding{177}%
  \or \ding{178}\or \ding{179}\or \ding{180}\or \ding{181}\else *\fi\relax}
\newcommand\mynumb[1]{\ifcase#1 \or \ding{182}\or \ding{183}\or
  \ding{184}\or \ding{185}\or \ding{186}\or \ding{187}%
  \or \ding{188}\or \ding{189}\or \ding{190}\or \ding{191}\else *\fi\relax}

\copyrightyear{2020}
\acmYear{2020}
\setcopyright{acmcopyright}
\acmConference[FPGA '20] {2020 ACM/SIGDA International Symposium on Field-Programmable Gate Arrays}{February 23--25, 2020}{Seaside, CA, USA}
\acmBooktitle{2020 ACM/SIGDA International Symposium on Field-Programmable Gate Arrays (FPGA'20), February 23--25, 2020, Seaside, CA, USA}
\acmPrice{15.00}
\acmDOI{10.1145/3373087.3375306}
\acmISBN{978-1-4503-7099-8/20/02}

\begin{document}

\fancyhead{}

\title{\textit{AutoDNNchip}: An Automated DNN Chip Predictor and Builder for Both FPGAs and ASICs}

\author{ Pengfei Xu$^1$, Xiaofan Zhang$^2$, Cong Hao$^2$, Yang Zhao$^1$, Yongan Zhang$^1$, Yue Wang$^1$, Chaojian Li$^1$, Zetong Guan$^1$, Deming Chen$^{2}$, Yingyan Lin$^1$} 
\affiliation{ 
	\institution{$^1$Rice University, TX, USA, $^2$University of Illinois at Urbana-Champaign, IL, USA}
	\textit{\{eiclab, zy34, yz87, yw68, cl114, zg20, yingyan.lin\}@rice.edu, \{xiaofan3, congh, dchen\}@illinois.edu}
}

\begin{abstract}
Recent breakthroughs in Deep Neural Networks (DNNs) have fueled a growing demand for domain-specific hardware accelerators (i.e., DNN chips). However, designing DNN chips is non-trivial because: (1) mainstream DNNs have millions of parameters and operations; (2) the design space is large due to the numerous design choices of dataflows, processing elements, memory hierarchy, etc.; and (3) an algorithm/hardware co-design is needed to allow the same DNN functionality to have a different decomposition, which would require different hardware IPs that correspond to dramatically different performance/energy/area tradeoffs. Therefore, DNN chips often take months to years to design and require a large team of cross-disciplinary experts. To enable fast and effective DNN chip design, we propose \textit{AutoDNNchip} $-$ a DNN chip generator that can automatically generate both FPGA- and ASIC-based DNN chip implementation (i.e., synthesizable RTL code with optimized algorithm-to-hardware mapping (i.e., dataflow)  ) given DNNs from machine learning frameworks (e.g., PyTorch) for a designated application and dataset without humans in the loop. Specifically, \textit{AutoDNNchip} consists of two integrated enablers: (1) a \textit{Chip Predictor}, built on top of a graph-based accelerator representation, which can accurately and efficiently predict a DNN accelerator's energy, throughput, latency, and area based on the DNN model parameters, hardware configuration, technology-based IPs, and platform constraints; and (2) a \textit{Chip Builder}, which can automatically explore the design space of DNN chips (including IP selection, block configuration, resource 
balance, etc.), optimize chip design via the \textit{Chip Predictor}, and then generate synthesizable RTL code with optimized dataflows to achieve the target design metrics. Experimental results show that our \textit{Chip Predictor}'s predicted performance differs from real-measured ones by $<$10\% when validated using 15 DNN models and 4 platforms (edge-FPGA/TPU/GPU and ASIC). Furthermore, both the FPGA- and ASIC-based DNN accelerators generated by our \textit{AutoDNNchip} can achieve better (up to 3.86$\times$ improvement) performance than that of expert-crafted state-of-the-art accelerators, showing the effectiveness of \textit{AutoDNNchip}. Our open-source code can be found at https://github.com/RICE-EIC/AutoDNNchip.git.

\end{abstract}
\vspace{-16pt}

\maketitle
{\fontsize{8pt}{8pt} \selectfont
\textbf{ACM Reference Format:}\\
Pengfei Xu, Xiaofan Zhang, Cong Hao, Yang Zhao, Yongan Zhang, Yue Wang, Chaojian Li, Zetong Guan, Deming Chen, Yingyan Lin. 2020. \textit{AutoDNNchip}: An Automated DNN Chip Predictor and Builder for Both FPGAs and ASICs. In \textit{2020 ACM/SIGDA International Symposium on Field-Programmable Gate Arrays (FPGA'20), February 23-25, 2020, Seaside, CA, USA.} ACM, New York, NY, USA, 11 pages. https://doi.org/10.1145/3373087.3375306 }

\vspace{-0.7em}
\section{Introduction}
\vspace{-0.1em}
We have seen the rapid adoption of Deep Neural Networks (DNNs) for solving real-life problems, such as image classification~\cite{VGG2014, wang2019e2}, object detection~\cite{ren2015faster}, natural language processing~\cite{xiong2017microsoft}, etc. Although DNNs enable high-quality inferences, they also require a large amount of computation and memory demand during deployment due to their inherently immense complexity \cite{liu2018adadeep,wang2018energynet,fracskip,deepkmeans,wang2019dual}. Moreover, DNN-based applications often require not only high inference accuracy, but also aggressive hardware performance, including high throughput, low end-to-end latency, and limited energy consumption. Recently, we have seen intensive studies on DNN accelerators in hardware, which attempt to take advantage of different hardware design styles, such as GPUs, FPGAs, and ASICs, to improve the speed and efficiency of DNN inference and training~\cite{zhang2015optimizing,7780065,mlfpga_liu2011real,liu2017throughput,mlfpga_zhang2017high,mlfpga_zhuge2018face,8050797,zhang2018dnnbuilder,TPU,edgetpu,8351679,du2015shidiannao,eyeriss}.

However, developing customized DNN accelerators presents significant challenges as it asks for cross-disciplinary knowledge in machine learning, micro-architecture, and physical chip design. Specifically, to build accelerators on FPGAs or ASICs, it is inevitable to include (1) customized architectures for running DNN workloads, (2) RTL programming for implementing accelerator prototypes, and (3) reiterative verifications for validating the functionality correctness. 
The whole task requires designers to have a deep understanding of both DNN algorithms and hardware design. In response to the intense demands and challenges of designing DNN accelerators, we have seen rapid development of high-level synthesis (HLS) design flow~\cite{vivado_HLS,hls_chen2005xpilot,hls_chen2009lopass,hls_rupnow2011high} and DNN design automation frameworks~\cite{wang2016deepburning, zhang2018caffeine, guan2017fp, venkatesanmagnet,wang2018design,zhang2018dnnbuilder} that improve the hardware design efficiency by allowing DNN accelerator design from high-level algorithmic descriptions and using pre-defined high-quality hardware IPs. Still, they either rely on hardware experts to trim down the large design space (e.g., use pre-defined/fixed architecture templates and explore other factors~\cite{venkatesanmagnet, zhang2018dnnbuilder}) or conduct merely limited design exploration and optimization, hindering the development of optimal DNN accelerators that can be deployed into various platforms.

To address the challenges above, we propose \textit{AutoDNNchip}, an end-to-end automation tool for generating optimized FPGA- and ASIC-based accelerators from machine learning frameworks (e.g., Pytorch/Tensorflow) and providing fast and accurate performance estimations of hardware accelerators implemented on various tar-

\noindent geted devices. The main contributions of this paper are as follows:
\begin{itemize}[leftmargin=*]
    \item \textbf{\textit{One-for-all Design Space Description}.} We make use of a graph-based representation that can unify design factors in all of the three design abstraction levels (including IP, architecture, and hardware-mapping levels) of DNN accelerator design, allowing highly flexible architecture configuration, scalable architecture/IP/mapping co-optimization, and algorithm-adaptive accelerator design. \\

\vspace{-1.3em}
    \item \textbf{\textit{Chip Predictor}.} Built on top of the above design space description, we propose a DNN \textit{Chip Predictor}, a multi-grained performance estimation/simulation tool, which includes a coarse-grained, analytical-model based mode and a fine-grained, run-time-simulation based mode. 
Experiments using 15 DNN models and 4 platforms (edge-FPGA/TPU/GPU and ASIC) show that our \textit{Chip Predictor}'s predicted error is within $10\%$ of real-measured energy/latency/resource-consumption. 
    \item \textbf{\textit{Chip Builder}.} We further propose a DNN \textit{Chip Builder}, which features a two-stage Design Space Exploration (DSE) methodology. Specifically, our \textit{Chip Builder} realizes: (1) an architecture/IP design based on the \textit{Chip Predictor}'s coarse-grained, analytical-model based prediction for a 1st-stage fast exploration and optimization, and (2) an IP/pipeline design based on the \textit{Chip Predictor}'s fine-grained, run-time-simulation based prediction as a 2nd-stage IP-pipeline co-optimization. Experiments show that the \textit{Chip Builder}'s 1st-stage DSE can efficiently rule out infeasible choices, while its 2nd-stage co-optimization can effectively boost the performance of remaining design candidates, e.g., 36.46\% throughput improvement and 2.4$\times$ idle cycles reduction.
    \item \textbf{\textit{AutoDNNchip}.} Integrating the aforementioned two enablers (i.e., \textit{Chip Predictor} and \textit{Chip Builder}), we develop \textit{AutoDNNchip}, which can automatically generate optimized DNN accelerator implementation (i.e., synthesizable RTL implementation) given the user-defined DNN models from machine learning frameworks (e.g., Pytorch), application-driven specifications (e.g., energy and latency), and resource budget (e.g., size of the processing array and memories). Experiments demonstrate that the optimized FPGA- and ASIC-based DNN accelerators generated by \textit{AutoDNNchip} outperform the recent award-winning design~\cite{zhang2019skynet} by 11\% and a state-of-the-art accelerator~\cite{mobile} by up to 3.86$\times$ .
    \vspace{-0.5em}
\end{itemize}


As an automated DNN accelerator design tool, \textit{AutoDNNchip} \textbf{is the first to highlight} all of the following features: (1) efficient and accurate performance prediction of DNN accelerators on 4 platforms, enabling fast optimal algorithm-to-accelerator mapping design and algorithm/accelerator co-design/co-optimization;
(2) a design space description that unifies the descriptions of design factors from all of the three design abstraction levels in DNN accelerators into one directed graph, supporting arbitrary accelerator architectures (e.g., both homogeneous and heterogeneous IPs and their inter-connections), and (3) can automatically generate both FPGA- and ASIC-based DNN accelerator implementation that outperforms expert-crafted state-of-the-art designs for various applications. 

\vspace{-0em}

\vspace{-0.5em}
\section{Background and related works}
\vspace{-0.0em}
\label{sec:background}

\begin{figure}[!t]
    \vspace{-1.0em}
    \centerline{\includegraphics[width=0.46\textwidth]{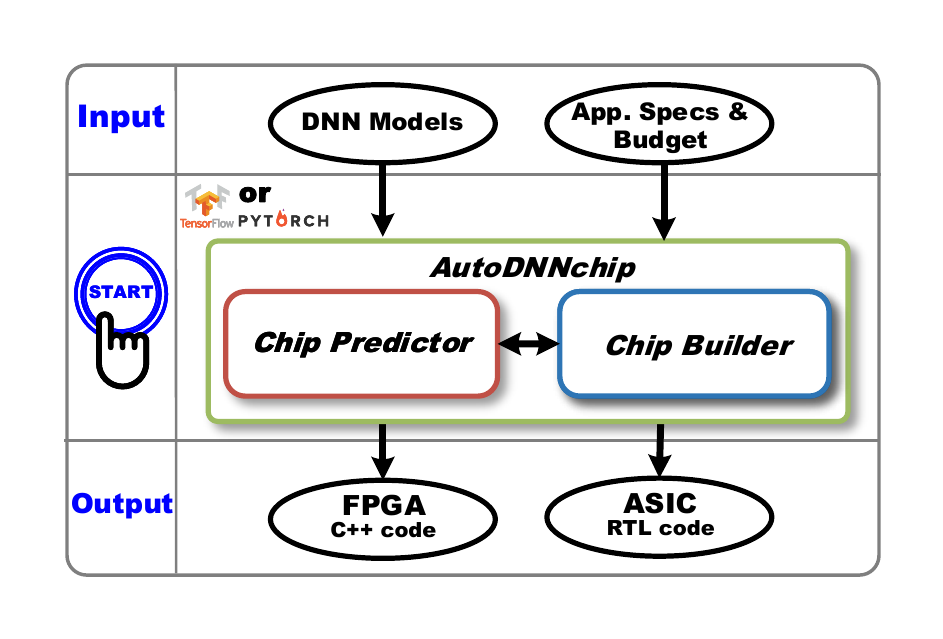}}
    \vspace{-2.6em}
  \caption{\textbf{Overview of the proposed \textit{AutoDNNchip} framework, which accepts user-defined DNN models/datasets and application-driven specifications to automatically generate optimized FPGA- or ASIC-based DNN accelerator designs.
    }}
    \vspace{-3.5em}
    \label{fig:autodnn-overview}
\end{figure}



\textbf{FPGA- and ASIC-based DNN Accelerators.}
There has been intensive study in customized  FPGA- and ASIC-based DNN accelerators. The accelerator in~\cite{zhang2015optimizing} uses loop tiling for accelerating convolutional layers on FPGAs. The  DNNBuilder accelerator~\cite{zhang2018dnnbuilder} applies an optimal resource allocation strategy, fine-grained layer-based pipeline, and column-based cache to deliver high-quality FPGA-based DNN accelerators.  The work in~\cite{liu2017throughput} proposes a throughput-oriented accelerator with multiple levels (i.e., task, layer, loop, and operator levels) of parallelisms. 
The recent designs in \cite{hao2019fpga,zhang2019skynet} introduce a hardware-efficient DNN and accelerator co-design strategy by considering both algorithm and hardware optimizations, using DNN building blocks (called Bundles) to capture hardware constraints.
For ASIC-based DNN accelerators, efforts have been made in both industry and academia, where representative ones include TPU~\cite{TPU,edgetpu}, ShiDianNao~\cite{du2015shidiannao}, and Eyeriss~\cite{eyeriss}, and different accelerators exploit different optimizations for various applications.


\textbf{DNN Accelerator Performance Prediction.}
For designing FPGA-based DNN accelerators, current practice usually relies on roofline models~\cite{zhang2015optimizing} or customized analytical tools~\cite{zhang2018dnnbuilder, liu2017throughput} to 
estimate the achievable performance. 
For ASIC-based accelerators, recently published designs~\cite{eyeriss, kwon2018maestro, parashar2019timeloop} introduce various performance prediction methods. Eyeriss \cite{eyeriss} proposes an energy model for capturing the energy overhead of the customized memory and computation units and a delay model that simplifies the latency calculation. Similarly, MAESTRO~\cite{kwon2018maestro} develops an energy estimation model that considers hardware design configurations and memory access behaviors, while Timeloop~\cite{parashar2019timeloop} adopts a loop-based description of targeted workloads and analyzes the data movement and memory access for latency estimation. 

\textbf{DNN Accelerator Generation.}
The tremendous need for developing FPGA-/ASIC-based DNN accelerators motivates the development of automated DNN accelerator generation. For example, DeepBurning \cite{wang2016deepburning} is a design automation tool for building FPGA-based DNN accelerators with customized design parameters using a pre-constructed RTL module library. DNNBuilder~\cite{zhang2018dnnbuilder} and FP-DNN~\cite{guan2017fp} propose end-to-end tools that can automatically generate optimized FPGA-based accelerators from high-level DNN symbolic descriptions in Caffe/Tensorflow frameworks. Caffeine \cite{zhang2018caffeine} is another automation tool that provides guidelines for choosing FPGA hardware parameters, such as the number of processing elements (PEs), bit precision of variables, and parallel data factors. By using these automation tools, it is easier to bridge the gap between fast DNN construction in popular machine learning frameworks and slow implementation of targeted hardware accelerators. 

\vspace{-0.2em}
\section{Overview of \textit{AutoDNNchip} } 
\vspace{-0.1em}
\label{sec:prop}
\begin{figure*}[!t]
\vspace{-3.0em}
    \centerline{\includegraphics[width=180mm]{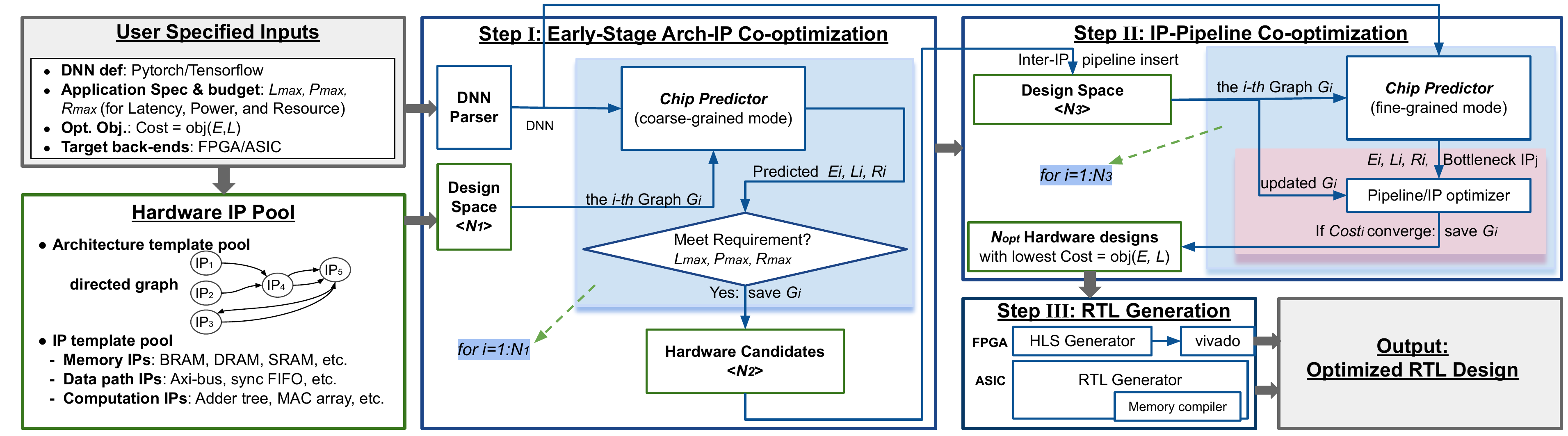}}
    \vspace{-1.2em}
    \caption{\textbf{\textit{AutoDNNchip}'s three-step design flow for the design space exploration, optimization, and DNN-to-RTL generation.}
    }
    \vspace{-1.2em}
    \label{fig:autodnn-sub}
\end{figure*}


Fig.~\ref{fig:autodnn-overview} shows an overview of the proposed \textit{AutoDNNchip}, which can automatically generate optimized FPGA- or ASIC-based DNN accelerators as well as an optimal algorithm-to-hardware mapping (i.e., dataflow), according to three customized inputs: (1) the high-level DNN descriptions trained in desired datasets, (2) the application-driven specifications regarding DNN inference quality and hardware performance, and (3) the available resources of targeted  platforms.  
The realization of \textit{AutoDNNchip} is achieved by the proposed \ul{\textit{One-for-all Design Space Description}} (see \textbf{Section~\ref{sec:graph}}), \ul{\textit{Chip Predictor}} (see \textbf{Section~\ref{sec:chip_pred}}), and \ul{\textit{Chip Builder}} (see \textbf{Section~\ref{sec:dse}}). 

\begin{table}[!b]
\vspace{-2.0em}
\caption{A summary of DNN accelerators' design factors.}
\vspace{-1.3em}
\def\arraystretch{1.2}
\centering
\scriptsize
\begin{tabular}{|c|c|c|c|}
\hline
{\textbf{Design factor}} & {\textbf{Description}} & {\textbf{Back-end}}  & {\textbf{Opt. level}} \\
\hline
{$B_{W} ,B_{A}, B_{Acc}$ $^a$} & {Bit precision} & {F, A $^b$} & {IP, Accuracy req.}\\
\hline
{$Freq.$} & {Clock frequency} & {F, A} & {Arch., IP}\\
\hline
{$Arch_{mem}$} & {Memory tech/hierarchy/volume} & {A} & {Arch., IP, Mapping}\\
\hline
{$Arch_{pe}$} & {PE array architecture} & {F, A} & {Arch., IP, Mapping}\\
\hline
{$Bw$} & {Port/Bus width for data transfer} & {A} & {Arch., IP}\\
\hline
{$Malloc$} & {Memory allocation} & {F, A} & {Arch., IP, Mapping}\\
\hline
{$Data$ $Schedule$} & {DNN to accelerator mapping} & {F, A} & {Arch., IP, Mapping}\\
\hline

\end{tabular}
\vspace{-0em}
\\
\begin{tablenotes}
\item{$^a$ $B_{W} ,B_{A}, B_{Acc}$: Bit precision for weights, activations, accumulations}\\
\item{$^b$ A: ASIC design; F: FPGA design}
\end{tablenotes}
\label{tab:design_space}
\vspace{-3.0em}
\end{table}

One of the major challenges that \textit{AutoDNNchip} needs to overcome is the lack of effective representations of DNN accelerators' large design space given the numerous design choices (e.g., dataflows, the number of pipeline stages, parallelism factors, memory hierarchy, etc.), as a precise and concise representation is a precondition of valid DNN accelerator design. To address this challenge, we propose a \textit{One-for-all Design Space Description}, which is an object-oriented graph-based definition for DNN accelerator design that can unify the description of design factors from all of the \textbf{three} design abstraction levels into \textbf{one} directed graph. Furthermore, \textit{AutoDNNchip} features another two key enablers, the \textit{Chip Predictor} and the \textit{Chip Builder}. Specifically, the proposed \textit{Chip Predictor} can accurately and efficiently estimate the energy, throughput, latency, and area overhead of DNN accelerators based on the parameters that can characterize the algorithms, hardware architectures, and technology-based IPs. The proposed \textit{Chip Builder} can automatically (1) explore the design space of DNN accelerators (including IP selection, block configuration, resource balance, etc.), (2) optimize chip designs via the \textit{Chip Predictor}, and (3) generate synthesizable Verilog code to achieve target design metrics. 

\begin{table}[h]
\vspace{0.5em}
\caption{A summary of attributes for the nodes and edges in the graph-based description}
\vspace{-1.2em}
\def\arraystretch{1.2}
\centering
\scriptsize
\begin{tabular}{|c|c|c|}
\hline
{\textbf{Compo.$^a$}} & {\textbf{Hardware meaning}} & {\textbf{Attributes}} \\
\hline
\multirow{3}{*}{Node} & {Memory IPs} & {Impl., Freq., Vol., Prec., Dt., StM., E, L$^b$}\\
\cline{2-3}
 & {Computation IPs} & {Impl., Freq., Prec., StM., E, L}\\
\cline{2-3}
 & {Data Path IPs} & {Impl., Freq., Bw.$^c$ Prec., Dt., StM., E, L}\\
\cline{2-3}
\hline
{Edge} & {IP inter-connections (IP dependency)} & {Start, End $^d$}\\
\hline
\end{tabular}
\begin{tablenotes}
\item{$^a$ Compo.: graph components including nodes and directed edges;}\\
\item{$^b$ Impl.: implementation, e.g., 14nm DRAM, 28nm SRAM, DSP48E, AXI-bus, sync FIFO, etc.;}\\
\item{$\quad$ Freq.: clock frequency (MHz); Vol.: volume/capacity (bits); Prec.: bit precision;}\\
\item{$\quad$ Dt.: data type including weights, input activations, and partial sums.; E/L: energy\&latency overhead; StM.: the state machine storing all the states (including needed inputs and generated outputs) through the whole execution process;}\\
\item{$^c$ Bw.: port/bus width; $^d$ Start \& End: the start and ending node for the directed edge}.\\
\end{tablenotes}
\vspace{-0pt}
\label{tab:nodes_edges}
\vspace{-2.5em}
\end{table}

\vspace{-0.5em}
\section{One-For-All Design Space Description}
\begin{figure}[!t]
\vspace{-0.8em}
    \centerline{\includegraphics[width=1.0\linewidth]{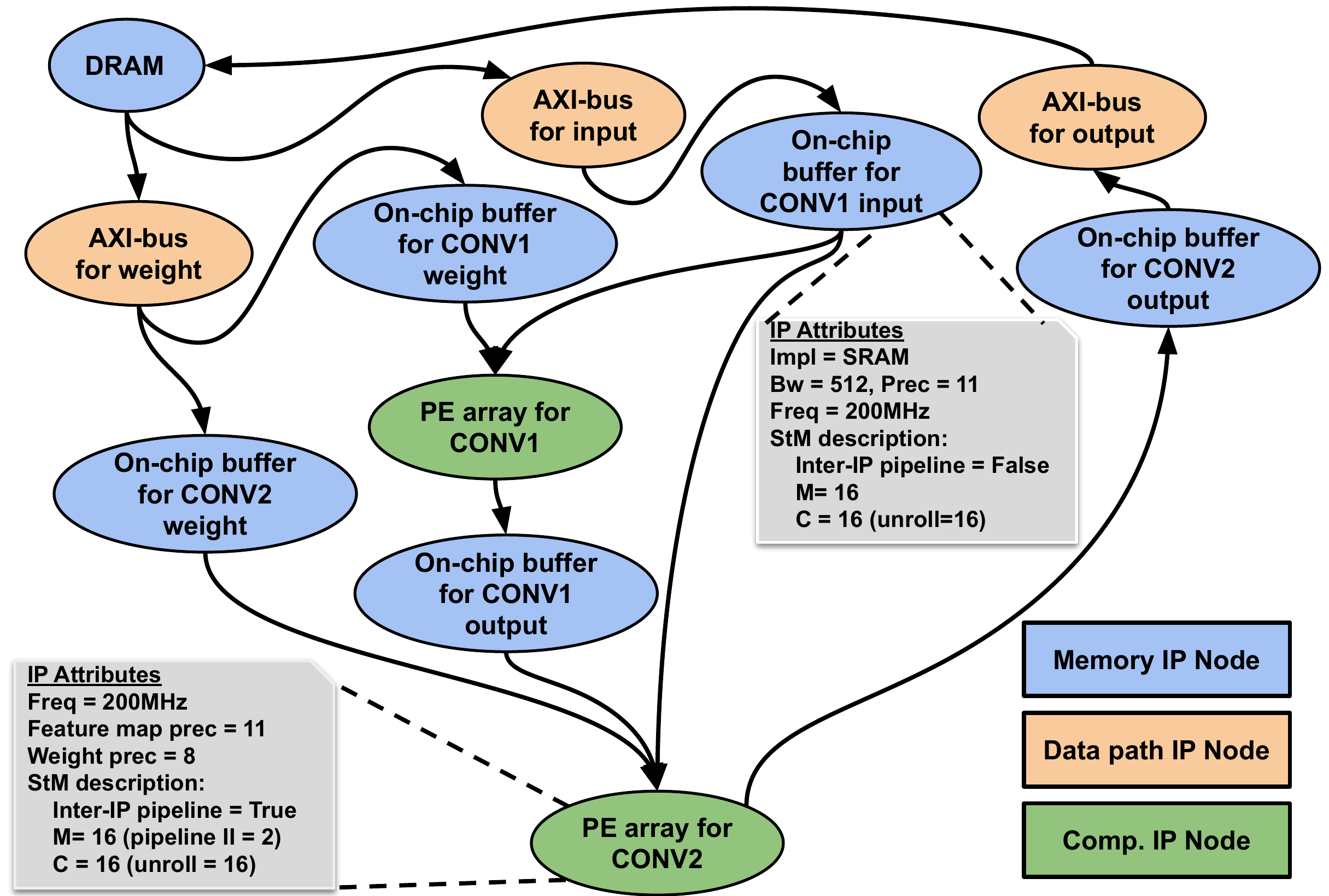}}
    \vspace{-0.8em}
    \caption{\textbf{An illustrative example of the graph-based design space description for a heterogeneous architecture to accelerate residual block in ResNet~\cite{he2016deep}, where M and C denote the output and input channel, respectively.}}
    \vspace{-0.7em}
    \label{fig:graph_arch}
\end{figure}

\vspace{-0.3em}
\label{sec:graph}
\textbf{Overview.}
It is well known that the design space of DNN accelerators can be very large. For effective and efficient design space exploration and optimization, it is critical that the design space can be precisely and concisely described, e.g., that the different design abstraction levels of optimization in DNN accelerator design, including architecture level, IP level, and hardware-mapping level, are considered. To this end, we adopt a \textit{One-for-all Design Space Description} that unifies the design factors of the three levels into one directed graph. Table~\ref{tab:design_space} lists the design factors which are sufficient for most cases and the last column shows the levels of design/optimization that may influence the corresponding factors. We can see that (1) most of the design factors are related to cross-level optimization which also reflects the fact that DNN accelerators have a large design space and (2) optimization at merely one level (or one hardware component) does not guarantee overall system performance. We thus adopt an object-oriented directed graph for the DNN accelerator design space description, an illustrative example of which is shown in Fig.~\ref{fig:graph_arch}. Specifically, a basic directed graph is first constructed using the PE array architecture, memory architecture and mapping/dataflow factors, where each node in the graph denotes a computation/data-path/memory IP and each directed edge denotes an inter-connection between nodes whose direction is determined by the corresponding data movement's direction. Proper attributes (e.g., those in Table~\ref{tab:nodes_edges}) are then assigned to the nodes and edges of the directed graph in an object-oriented manner. In the following subsections, we will briefly describe four graph-based accelerator templates corresponding to four state-of-the-art DNN accelerators which are stored in the \textit{Hardware IP Pool} (see Fig.~\ref{fig:autodnn-sub} under the \textit{User Specified Inputs}) of  \textit{AutoDNNchip} together with other templates to provide a sufficient number of design candidates, and then discuss the IP attributes for the nodes and edges.

\begin{figure}[!b]
\vspace{-0.5em}
    \centerline{\includegraphics[width=1\linewidth]{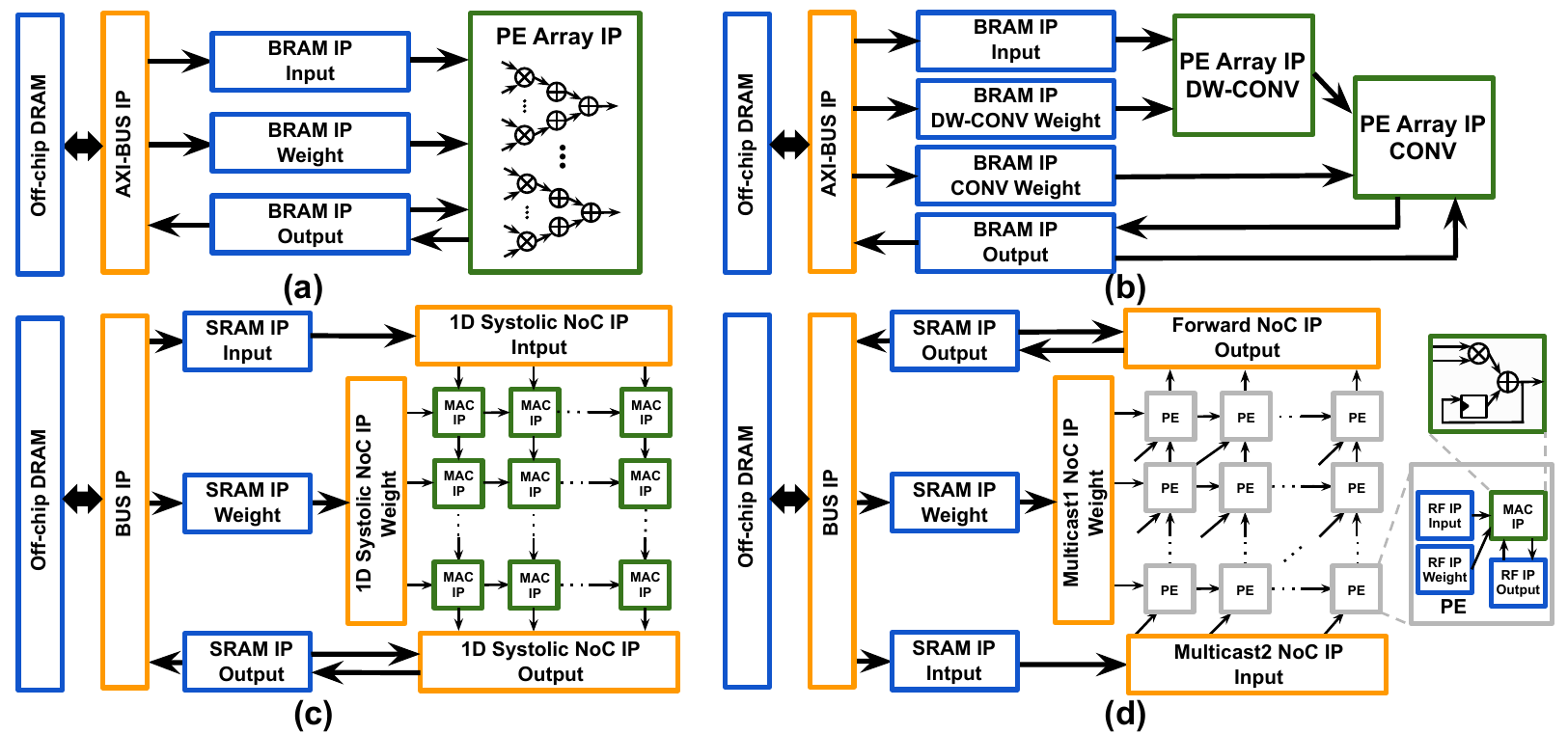}}
    \vspace{-1em}
    \caption{\textbf{An illustration of 4 architecture templates in our \textit{Hardware IP Pool} including 2 architectures for both state-of-the-art FPGA- and ASIC-based DNN accelerators.}}
    \vspace{-0.5em}
    \label{fig:arch_temp}
\end{figure}

\textbf{Graph-based Accelerator Templates.}
Fig.~\ref{fig:arch_temp} shows four graph-based accelerator template examples for describing DNN accelerators that can be translated into real hardware implementation by applying appropriate IP attributes. Specifically,
Fig.~\ref{fig:arch_temp} (a) shows a spatial architecture based on a single adder-tree based computation IP, which is a commonly-used architecture on FPGA-based accelerators;
Fig.~\ref{fig:arch_temp} (b) is a graph with 2 different computation IPs, including depth-wise convolutional (denoted as DW$\_$CONV) and normal convolutional (denoted as CONV) ones commonly adopted in compact DNN models, and two BRAM IPs that handle the memory data arrangement for the computation IPs; 
Fig.~\ref{fig:arch_temp} (c) is an architecture template for TPU~\cite{TPU} type DNN accelerators using a systolic array; and Fig.~\ref{fig:arch_temp} (d) shows the 
graph-based representation for DNN accelerators with Eyeriss~\cite{eyeriss} type architectures, where the data path IPs (i.e., NoC IPs in Fig.~\ref{fig:arch_temp} (d)) between PEs describe the local data reuse patterns of inputs, outputs, and weights.

\begin{figure}[!t]
    \centerline{\includegraphics[width=1\linewidth]{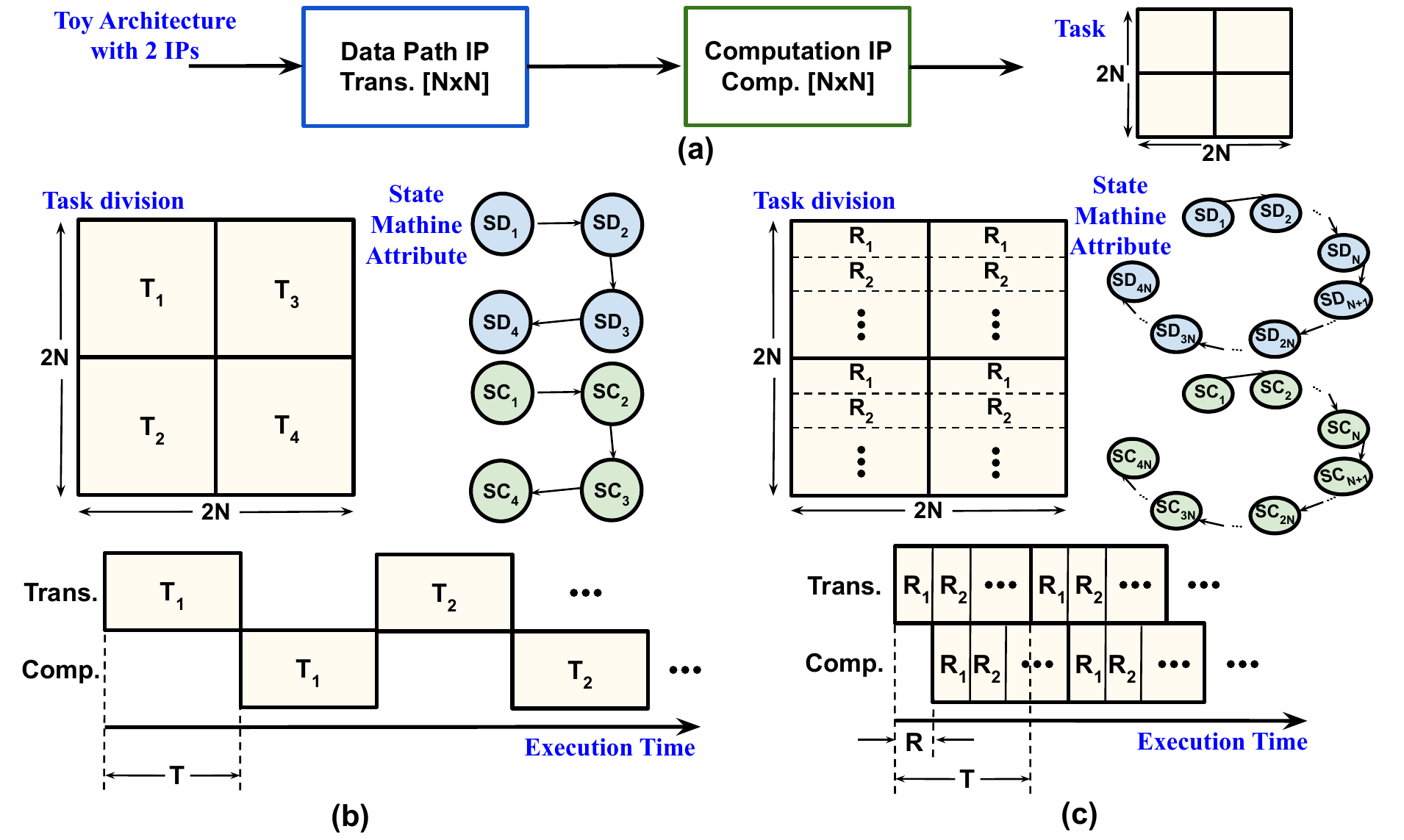}}
    \vspace{-1.5em}
    \caption{\textbf{A toy example of IP's state machine attribute w/o and w/ considering inter-IP pipeline effects:
    (a) a simple architecture with 2 IPs, i.e., one data path IP and one computation IP; the task division, state machine, and the run-time process when (b) excluding the inter-IP pipeline and (c) considering the inter-IP pipeline, where SD and SC denote the state for data path IP and computation IP, respectively. }}
    \vspace{-1.5em}
    \label{fig:toy2}
\end{figure}

\textbf{IP Attributes.} Table~\ref{tab:nodes_edges} summarizes the attributes for three types of node IP including memory (e.g., BRAM and off-Chip DRAM ), data access (e.g., bus), and computation hardware that characterizes the corresponding design, as elaborated below:
(1) The \textit{Implementation or Impl.} attribute refers to the required hardware resource for implementing the IPs, e.g., DRAM and SRAM for implementing memory IPs, and AXI-bus and NoC for implementing data path IPs; (2) The \textit{state machine or StM.} attribute is used to describe when the IPs will update their states between computation and loading/unloading data, where each state defines both the needed input address and generated output address.  Fig.~\ref{fig:toy2} shows that different pipeline designs can be captured by the IPs' state machine attribute:  Fig.~\ref{fig:toy2} (b) and (c) illustrate two kinds of designs (w/o and w/ inter-IP pipeline) and their corresponding state machine definition, respectively, where there are more states in Fig.~\ref{fig:toy2} (c) for capturing the inter-IP pipeline between data transfer and computation IPs; (3) The \textit{data precision or Prec.} attribute refers to the IPs' bit precision; (4) The \textit{clock Frequency or Freq.} and \textit{energy/latency or E/L} attributes capture the operating clock frequency and required energy/latency for IPs; and (5) The \textit{port/bus width or Bw} and \textit{memory volume or Vol.} attributes refers to the port/bus width of data path IPs and memory volume for memory IPs, respectively.

\vspace{-0.2em}
\section{The Proposed \textit{Chip Predictor}}
\label{sec:chip_pred}
\vspace{-0.2em}
\subsection{Overview}
\vspace{-0.2em}

\begin{figure}[!b]
    \vspace{-1em}
    \centerline{\includegraphics[width=0.9\linewidth]{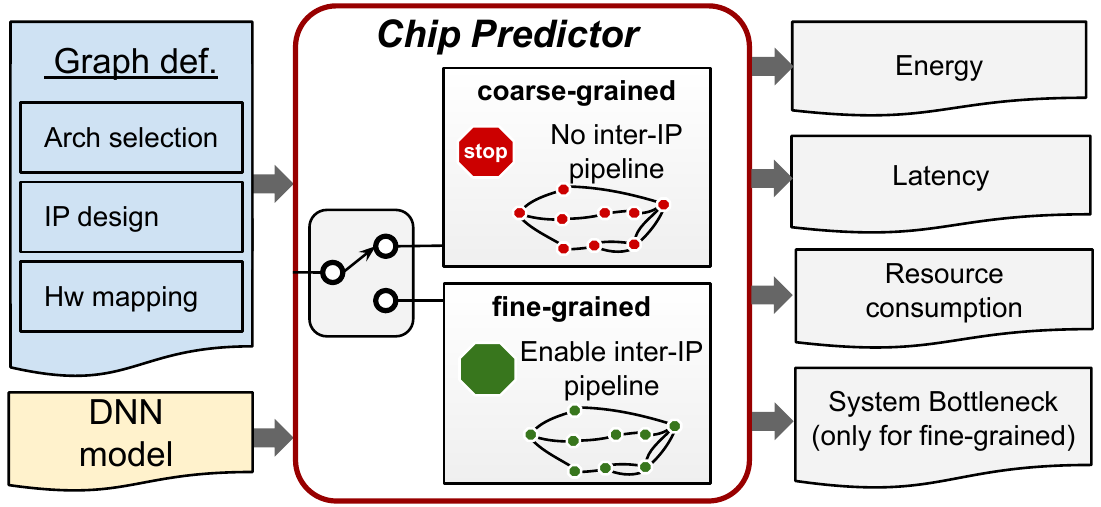}}
    \vspace{-1.2em}
    \caption{\textbf{An overview of the proposed \textit{Chip Predictor}.}}
    \vspace{-0.5em}
    \label{fig:chip_predictor}
\end{figure}

As shown in Fig.~\ref{fig:chip_predictor}, the proposed \textit{Chip Predictor} accepts DNN mod-

\noindent els (e.g., number of layers, layer structure, precision, etc.), hardware architectures (e.g., memory hierarchy, number of PEs, NoC design, etc.), hardware mapping, and IP design (e.g., unit energy/delay cost of a multiply-and-accumulate (MAC) operation and memory accesses to various memory hierarchies), and then outputs the estimated energy consumption, latency, and resource consumption when executing the DNN in the target accelerator defined by the given hardware architecture and hardware mapping. First, to capture the large search space and consider all the design abstraction levels (including architecture, IP, and hardware mapping levels), we construct a graph-based description that serves as one input of \textit{Chip Predictor}. Second, to match the different tradeoff requirements of the \textit{Chip Builder}'s two-stage DSE, which aims for efficient and accurate design space exploration and optimization, our \textit{Chip Predictor} adopts a mixed-granularity prediction: (1) a coarse-grained mode that can quickly provide IP performance estimation to enable the identification of critical paths when the inter-IP pipeline is not considered, in order to be used for the \textit{Chip Builder}'s early-stage architecture and IP exploration and selection; and (2) a fine-grained mode that can perform accurate performance prediction by considering the pipeline dependency between IPs based on run-time simulations, in order to be used for the \textit{Chip Builder}'s 2nd-stage DSE that targets IP-pipeline co-optimization.


\begin{figure}[!t]
    \vspace{-1.7em}
    \centerline{\includegraphics[width=1.0\linewidth]{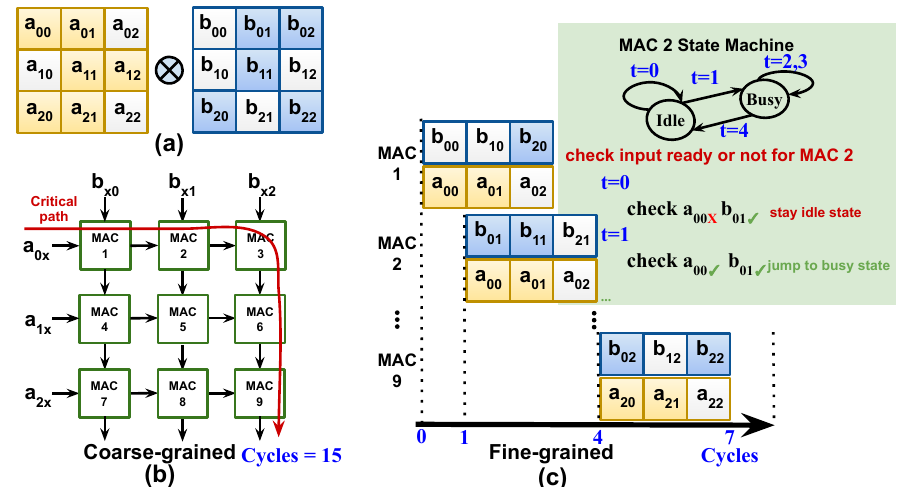}}
    \vspace{-1.3em}
    \caption{\textbf{A toy example using a systolic array, illustrating: (a) the corresponding matrix-matrix multiplication, and (b) coarse-grained and (c) fine-grained latency estimation.}}
    \vspace{-1.5em}           
    \label{fig:toy1}
\end{figure}

\vspace{-0.7em}
\subsection {The \textit{Chip Predictor}'s Coarse-grained Mode}

\label{subsec:coarse_pred}

\textbf{Overview.} The \textit{Chip Predictor}'s coarse-grained mode is analytical-model-based, i.e., using equations to formulate the accelerators' energy, critical path latency, and resource consumption given a DNN model and a graph-based hardware design description (see Fig.~\ref{fig:chip_predictor}). Specifically, the energy and latency of IPs are first calculated using: (1) analytical equations as described below and (2) the attributes of each IP, where the unit energy/latency costs are obtained from single-IP RTL implementation or simulations; and the energy and latency consumption of the whole DNN accelerator are formulated by considering: (1) the total energy and latency of all IPs when executing the DNN model and (2) the energy and latency overhead of the CPU and on-chip controller.

\textbf{Analytical-model-based Intra-IP Modeling.} If we define the energy, latency and resource utilization as $E$, $L$, and $R$, respectively, and use $ip_{comp}$, $ip_{dp}$, $ip_{mem}$ to denote the computation IP, data path IP and the memory IP, respectively, the energy and latency of the computation IPs can be formulated by:
\setlength{\belowdisplayskip}{0pt} \setlength{\belowdisplayshortskip}{0pt}
\setlength{\abovedisplayskip}{0pt} \setlength{\abovedisplayshortskip}{0pt}
\begin{align}
    E_{ip_{comp}} &= e_1 + (\# states) \times (e_2 + e_{mac} \times U)\\
    \vspace{-0.3em}
    L_{ip_{comp}} &= l_1 + (\# states) \times l_{mac}
    \label{eq:e_comp}
\end{align}
where $(\# states)$ denotes the total number of the states in the IP state machine; $U$ denotes the unrolling factor (PE parallelism) for the computation IP; $e_{mac}$ and $l_{mac}$ denote the unit energy and latency costs for a MAC operation, respectively; $e_1$ and $l_1$ are the energy and latency overhead for warming up, i.e., configure the data path and pre-load data; and $e_2$ denotes the energy overhead of run-time control of CPU or on-chip logic units. Meanwhile,
the energy and latency of the data path IPs can be formulated by:
\begin{align}
    E_{ip_{dp}} &= e_3 + (\# states) \times (e_4 + V \times e_{bit})\\
    \vspace{-0.5em}
    L_{ip_{dp}} &= l_2 + (\# states) \times (l_3 + \frac{V}{P_w} \times l_{bit})
    \label{eq:e_dp}  
\end{align}
where $V$ denotes the total data volume (bits) needed to be trans-

\noindent ferred when the IPs are called; $P_w$ denotes the port width for the corresponding data path; $e_{bit}$ and $l_{bit}$ denote the unit energy and latency costs for each bit of data access, respectively; $e_3$ and $l_2$ are the energy and latency overhead for warming up, respectively; and $e_4$ and $l_3$ denote the energy and latency overhead of the run-time control of CPU or on-chip logic units, respectively.


\textbf{Analytical-model-based Inter-IP Modeling.}
For the system performance, including energy, latency, and resource consumption of a convolutional layer or a DNN building block (e.g., the Bundle in \cite{zhang2019skynet,hao2019fpga}), the resource and energy consumption are obtained by summing up that of all the IPs in the graph, and the total latency can be calculated by summing up the latency of all the IPs on the critical path of the graph, i.e.,
\vspace{1.0em}
\begin{align}
    &R_{mem} = \sum_{ip_{mem} \in G} Vol_{ip_{mem}}\\
    &R_{mul} = \sum_{ip_{comp} \in G} U_{ip_{comp}} + R_{mul_{dec}}\\
    &E = \sum_{ip\in G} E_{ip}\\
    &L = \max_{path \in G} \sum_{ip \in path} L_{ip}
\label{eq:inter-ip}   
\end{align}
where $G$ denotes the whole graph; $R_{mem}$ denotes the total memory volume consumption for one type of memory; $R_{mul}$ denotes the total number of multipliers used in both the computation IPs and when decoding the memory address, with the latter denoted as $R_{mul_{dec}}$.
Regarding the latency estimation, the inter-IP pipeline effects are excluded in the coarse-grained mode and it can be captured in the fine-grained mode of \textit{Chip Predictor} (see Section~\ref{subsec:fine_pred}). As a toy example, Fig.~\ref{fig:toy1} (b) and (c) illustrate the latency estimation when operating a matrix-matrix multiplication in a systolic array using both the coarse-grained mode and fine-grained mode, where the resulting estimated latency results are 15 and 7 cycles, respectively.
\vspace{-3.0em}
\subsection{The \textit{Chip Predictor}'s Fine-grained Mode}\label{subsec:fine_pred}
 \vspace{-0.0em}
\textbf{Overview.}
In the fine-grained mode of the \textit{Chip Predictor}, we adopt: (1) \textit{Algorithm}~\ref{alg:sim} to perform run-time simulations based on inter-IP pipeline to obtain the corresponding inter-IP latency; and (2) the \textit{Chip Predictor}'s coarse-grained mode to get the IPs' energy and latency for estimating the intra-IP performance.

\textbf{Implementation.}
The run-time simulation algorithm is described in \textit{Algorithm}~\ref{alg:sim}, where each IP (denoted as $ip$) has (1) its neighbour IPs on the graph defined as $ip.prev$ and $ip.next$, respectively, and $ip$ will use the data from $ip.prev$ as its inputs and pass its outputs to $ip.next$; and
(2) a state machine to store different states (including its needed inputs and generated outputs) through the whole execution process. 
For each clock cycle in the simulation, $ip$ can jump to the next state when (1) it has finished generating all the outputs in its current state (i.e., $ip$ is in an idle status) and (2) $ip.prev$ has generated all the inputs $ip$ needed for the next state. 
If $ip$ is in an idle status but its needed inputs are not ready from $ip.prev$, it will continue to wait on the idle status, resulting in an increase of the idle cycles associated with this IP;
If $ip$ is in a busy status, it will generate its outputs and jump to an idle status when it finishes generating all the outputs in this state.

For better understanding, Fig.~\ref{fig:toy1} uses a toy example to show that the~\textit{Chip Predictor}'s fine-grained mode (see Fig.~\ref{fig:toy1} (c)) can more accurately estimate the required latency than its coarse-grained mode. In this 3$\times$3 systolic array with the local-data-forwarding and computation operations being pipelined, we assume each MAC unit takes 3 cycles to do the computation and 1 cycle to forward the data to its nearby MAC units. In the coarse-grained mode case, we add the intra-IP latency in the graph's critical path to estimate the overall latency (see Fig.~\ref{fig:toy1} (b)), resulting in an estimated latency of 15 cycles. In the fine-grained mode case (see Fig.~\ref{fig:toy1} (c)), we define the state machine for each MAC unit and adopt \textit{Algorithm} \ref{alg:sim} to keep track of when each MAC unit jumps to the next state. In this particular example, MAC 2 will wait at cycle 0 since its required input data a00 is not ready, and it will jump to next state to start computing at cycle 1 when all its required inputs are ready. We can see that the fine-grained mode's estimated latency (7 cycles, the same as the ground truth) is more accurate for modeling the overlapped computation and data transferring in this example. In practical designs, the overall latency is not determined by merely one stage, so the \textit{Chip Builder} will launch the \textit{Chip Predictor} to simulate the whole graph iteratively in order to generate an optimal design for the whole accelerator system.

\begin{figure}
\vspace{-12pt}
    \begin{minipage}{0.47\textwidth}
        \begin{algorithm}[H]
            \caption{Run-time sim. in the fine-grained \textit{Chip Predictor}}
            \label{alg:sim}
            \begin{algorithmic}[1]
                \footnotesize
                \STATE{Input: One accelerator design described by graph $G$;}
                \STATE{\textbf{For} each $edge $ in $G$}
                    \STATE{\hspace{10pt} $ip_{start}\longleftarrow edge'$s starting node;}
                    \STATE{\hspace{10pt} $ip_{end}\longleftarrow edge'$s ending node;}
                    \STATE{\hspace{10pt} Add $ip_{start}$ to $ip_{end}.prev$;}
                    \STATE{\hspace{10pt} Add $ip_{end}$ to $ip_{start}.next$;}
                \STATE{Initialize energy and latency: $E =0$, $cycles=0$;}
                \STATE{\textbf{While} not all inference outputs are stored back}
                \STATE{\hspace{6pt} $cycles\longleftarrow cycles+1$;}
                \STATE {\hspace{6pt} \textbf{For} each $ip$ in $G$}
                \STATE {\hspace{15pt} \textbf{If} ($ip$ is $idle$) \& (all needed inputs $\in$ outputs of $ip.prev$)}
                \STATE {\hspace{22pt} $ip\longleftarrow busy$;}        
                \STATE {\hspace{22pt} $ip$ jumps to the next state;}    
                \STATE {\hspace{15pt} \textbf{If} ($ip$ is $idle$) \& (not all needed inputs $\in$ outputs of $ip.prev$)}
                \STATE {\hspace{22pt} $ip.idle\_cycles\longleftarrow ip.idle\_cycles+1$;}          
                \STATE {\hspace{15pt} \textbf{If} ($ip$ is $busy$) \& (not all outputs for $ip$ is ready)}
                \STATE {\hspace{22pt} Update the ready outputs for $ip$;}
                \STATE {\hspace{15pt} \textbf{If} ($ip$ is $busy$) \& (all outputs for $ip$ is ready)}
                \STATE {\hspace{22pt} $ip\longleftarrow idle$;}
                \STATE {\hspace{22pt} $E\longleftarrow E+E_{ip}$;}
                \STATE{$L\longleftarrow\frac{cycles}{{global}~{clk}~ {freq}}$;}
                \STATE{$ip_{bottleneck}\longleftarrow ip$ with minimum idle cycles.}
            \end{algorithmic}
        \end{algorithm}
    \end{minipage}
\vspace{-2.0em}
\end{figure}

 \vspace{-0.7em}
\section{The Proposed \textit{Chip Builder}}
\label{sec:dse}
 \vspace{-0.3em}
Fig.~\ref{fig:autodnn-sub} elaborates the design flow of \textit{AutoDNNchip} that leverages the \textit{Chip Builder}'s two-stage DSE engine. 
To effectively explore the design space (e.g., the design factors in Table~\ref{tab:design_space}), 
\textit{AutoDNNchip} involves three major steps as shown in Fig.~\ref{fig:autodnn-sub}:
(1) the 1st-stage DSE: an early stage architecture and IP configuration exploration to efficiently rule out infeasible designs using the \textit{Chip Predictor}'s coarse-grained mode;
(2) the 2nd-stage DSE: an inter-IP pipeline exploration and IP optimization to effectively boost the performance of the remaining design candidates resulting from the 1st-stage DSE;
and (3) a design validation through RTL generation and execution.

\textbf{Step I. Early Stage Architecture and IP Configuration Exploration.} As shown in the middle part of Fig.~\ref{fig:autodnn-sub}, this step considers the following exploration. \ul{First}, the DNN model from a mainstream machine learing framework is applied to the DNN parser to extract the DNN layer information, e.g., layer types (CONV, Pooling, ReLU, Reorg \cite{zhang2019skynet}, etc.), feature map inter-connections (Concat, Add, etc.), and layer shapes (shape of weight and feature map tensors). \ul{Second}, according to the given DNN model, performance requirements (e.g., latency and throughput) and hardware budgets (e.g., resource and power budget of FPGA or ASIC), a design space of size $N_1$ is generated by fetching commonly-used or promising 
hardware architecture templates and hardware IP templates from the \textit{Hardware IP pool}.
For example, when the given resource budgets are tight, a folded hardware architecture will be chosen instead of a flattened one; whereas flattened structures which facilitate IP pipelines are preferred when there are sufficient budgets. 
\ul{Third}, an architecture and IP configuration optimization is then performed to rule out most of the infeasible choices and trim down the design space to $N_2$ ($N_2<N_1$) promising candidates, e.g, more efficient with a lower latency. This fast early exploration makes use of the analytical nature of the \textit{Chip Predictor}'s coarse-grained mode.

\begin{figure}
\vspace{-12pt}
    \begin{minipage}{0.47\textwidth}
        \begin{algorithm}[H]
            \caption{IP-pipeline co-optimizationusing the \textit{Chip Builder}}
            \label{alg:second}
            \begin{algorithmic}[1]
                \footnotesize
                \STATE{Input: Design space $D_G$ with $N_2$ graphs;}
                \STATE{\textbf{For} each $G$ in $D_G$}
                    \STATE{\hspace{10pt}\textbf{For} each $edge $ in $G$}
                        \STATE{\hspace{18pt} $ip_{start}\longleftarrow edge'$s starting node;}
                        \STATE{\hspace{18pt} $ip_{end}\longleftarrow edge'$s ending node;}
                        \STATE{\hspace{18pt} Add $ip_{start}$ to $ip_{end}.prev$;}
                        \STATE{\hspace{18pt} Add $ip_{end}$ to $ip_{start}.next$;}
                    \STATE{\hspace{10pt} \textbf{While} simulated (using Algorithm 1) latency $L_G$ does not converge}
                    \STATE{\hspace{18pt} $ip \longleftarrow $ simulated bottleneck IP (i.e., $ip_{bottleneck}$ from Algorithm 1);}
                    \STATE{\hspace{18pt} \textbf{If} inter-IP pipeline is adopted for $ip$ and $ip.next$} \STATE{\hspace{26pt} allocate more resource to $ip$;}                     \STATE{\hspace{18pt} \textbf{Else}}
                    \STATE{\hspace{26pt} adopt inter-IP pipeline between $ip$ and $ip.next$;}
                    \STATE{\hspace{26pt} update the state machine of $ip$;}
                    \STATE{\hspace{26pt} update the state machine of $ip.next$;}
            \STATE{Select top $N_{opt}$ candidates in $D_G$}
            \end{algorithmic}
        \end{algorithm}
    \end{minipage}
\vspace{-2.0em}
\end{figure}

\textbf{Step II. Inter-IP Pipeline Exploration and IP Optimization.}
This step accepts the resulting $N_2$ designs and performs further exploration and IP optimization using \textit{Algorithm}~\ref{alg:second}. 
\ul{First}, inter-IP pipelines are inserted into different locations of the corresponding computation graphs, resulting in a new design space of size $N_3$, i.e., $N_3$ new graphs with different inter-IP pipeline designs. \ul{Second}, for each of these graphs, the bottleneck IPs will be recorded during \textit{Algorithm}~\ref{alg:sim}'s run-time simulations and then optimized via deeper inter-IP pipeline design or re-allocating more resource until convergence based on the \textit{Chip Predictor}'s fine-grained mode's predicted performance, as shown in \textit{Algorithm}~\ref{alg:second}.
\ul{Third}, the top $N_{opt}$ design candidates will be chosen according to the \textit{Chip Predictor}'s predicted energy consumption or/and latency, and then passed to the next step for validation through RTL generation and execution.

\textbf{Step III. Design Validation through RTL Generation and Execution.}
In this step, we generate RTL code for the top $N_{opt}$ optimized designs through an automated code generation procedure: 
(1) For the FPGA back-end, the generated files include the testbench for a board-level implementation, the binary file for the quantized-and-reordered weights, and the C-code for the HLS IP implementation. We use Vivado \cite{vivado_HLS} to actually generate the bitstream and meanwhile eliminate the designs that fail in place and route (PnR) to guarantee that \textit{AutoDNNchip}'s generated designs are valid;
(2) For the ASIC back-end, the generated files include the RTL testbench for the DNN model, the quantized-and-reordered weights, the synthesizable RTL code, and the memory specifications. The RTL code could be further passed to an EDA tool like Design Compiler and IC Compiler to generate gate-level/layout netlist, during which Memory Compilers could take the memory specifications to generate the memory design. 
After this step, all the output designs are fully validated with correct functionality.

\vspace{-0.2em}
\section{Experiment Results}
In this section, we evaluate the proposed \textit{AutoDNNChip} on 20 DNN models across 4 platforms (3 edge devices including edge-FPGA/TPU/GPU and 2 ASIC-based accelerators).


\vspace{-0.7em}
\subsection{Validation of the \textit{Chip Predictor}}
\vspace{-0.3em}
\label{subsec:val}

\textbf{Methodology and Setup.} Table~\ref{tab:val_setting} summarizes the details of our validation experiments for the \textit{Chip Predictor}, including the platforms, performance metrics, DNN models, methods to obtain the unit parameters, employed precision for the weights and activations, and frequency of the corresponding computation core.
\textbf{Methodology.} 
In order to conduct a solid validation, we validate the \textit{Chip Predictor} by comparing its predicted performance with actual device-measured ones on \ul{3 edge devices} (Ultra96 FPGA~\cite{ultra96}, edge TPU~\cite{edgetpu}, Jetson TX2~\cite{edgegpu}) and paper-reported ones of \ul{2 published ASIC-based accelerators} (Eyeriss~\cite{eyeriss} and ShiDianNao~\cite{du2015shidiannao}), when adopting the same experiment settings (e.g., clock frequency, DNN model and dataset, bit precision, architecture design, and dataflow, etc). \textbf{Benchmark DNN Models and Datasets.} For the 3 edge devices, we consider 15 representative compact/light-weight DNN models (see Table~\ref{tab:dnn_sk} and Table~\ref{tab:dnn_mb}, where the models in Table~\ref{tab:dnn_mb} use the ImageNet dataset~\cite{Deng09imagenet} and the models in Table~\ref{tab:dnn_sk} use the dataset in the System Design Contest of the DAC 2019 conference ~\cite{dac-contest}); for the 2 published DNN accelerators, we use the same benchmark models and datasets as the original papers.
\textbf{Unit Parameters.} The unit energy/latency parameters are obtained through either real-device measurement or synthesized RTL implementation as mentioned in Section~\ref{sec:chip_pred}. For the 3 edge devices, we measure the unit energy and latency by running the basic IP operations (such as the memory accesses and the MAC computation) over multiple sets of experiments under different settings and average the energy and latency values to get unit parameters. Specially, for memory accesses, we change the clock frequency, memory volume, port width, bit precision, and burst read length; for the MAC operations, the clock frequency, total number of MACs and parallelism of MACs are changed.
For the ASIC-based accelerators, the unit parameters are obtained either from the paper~\cite{eyeriss} or gate-level simulations of the synthesized RTL implementation on the same CMOS technology.



\begin{scriptsize}
\begin{table}[h!]
\vspace{-0.5em}
\caption{Experiment settings for the \textit{Chip Predictor}'s cross-platform/model/design/dataset validation.}
\vspace{-1.0em}
\def\arraystretch{1.2}
\centering
\scriptsize
\begin{tabular}{|c|c|c|c|c|c|c|}
\hline
\multirow{2}{*}{\textbf{Arch/Device}} & \multirow{2}{*}{\textbf{Metrics$^a$}} & \multirow{2}{*}{\textbf{DNNs}$^b$} & {\textbf{Unit}}& {\textbf{Precision}} & {\textbf{Freq.$^e$}}\\
 & ~ & ~ & {\textbf{Param.$^c$}}& {\textbf{<W,A>}$^d$} & {\textbf{(MHz)}}\\
\hline
{\textbf{Ultra96}~\cite{ultra96}} & {\textit{E, L, R}} & {Compact} & {Measured}& {<11, 9>} & {220}\\
\hline
{\textbf{Edge TPU}~\cite{edgetpu}} & {\textit{E, L}} & {Compact} & {Measured} & {<8, 8>} & {500}\\
\hline
{\textbf{Jetson TX2}~\cite{edgegpu}} & {\textit{E, L}} & {Compact} & {Measured} & {<32, 32>} & {1300}\\
\hline
{\textbf{Eyeriss}~\cite{eyeriss}} & {\textit{E, L, R}} & {AlexNet}  & {Reported} & {<16, 16>} & {250}\\
\hline
{\textbf{ShiDianNao}~\cite{du2015shidiannao}} & {\textit{E}} & {Small} & {Synthesized} & {<16, 16>}& {1000}\\
\hline
\end{tabular}
\\
\begin{tablenotes}
\item{$^a$ Metrics -- \textit{E}: energy, \textit{L}: latency, \textit{R}: resource;}
\item{$b$ DNN benchmarks -- Compact: see the 15 compact DNN models in Table~\ref{tab:dnn_sk} and Table~\ref{tab:dnn_mb}; AlexNet~\cite{NIPS2012_4824}; and Small: DNNs used in ~\cite{du2015shidiannao} (< 5 convolutional/fully-connected layers);}
\item{$c$ Methods to obtain the unit parameters;}
\item{$d$ Bit precision for different types of data, i.e., <weight precision, activation precision>;
}
\item{$e$ Clock frequency for the computation core.}
\end{tablenotes}
\label{tab:val_setting}
\vspace{-1.0em}
\end{table}
\end{scriptsize}

\begin{table}[h!]
\caption{The 10 model variants of the SkyNet backbone~\cite{zhang2019skynet}.}
\vspace{-1.0em}
\def\arraystretch{1.2}
\centering
\scriptsize
\begin{tabular}{|m{1.0cm}|m{0.3cm}|m{0.3cm}|m{0.3cm}|m{0.3cm}|m{0.3cm}|m{0.3cm}|m{0.3cm}|m{0.3cm}|m{0.3cm}|m{0.3cm}|}
\hline
{\textbf{DNN}} & {\textbf{SK}} & {\textbf{SK1}} & {\textbf{SK2}} & {\textbf{SK3}} & {\textbf{SK4}} & {\textbf{SK5}} & {\textbf{SK6}} & {\textbf{SK7}} & {\textbf{SK8}} & {\textbf{SK9}} \\
\hline
{\textbf{Size (MB)}} & {1.75} & {1.79} & {2.11} & {1.18} & {1.77} & {3.21} & {3.79} & {3.05} & {0.96} & {1.95} \\
\hline
{\textbf{Layer \#}} & {14} & {14} & {14} & {14} & {17} & {14} & {16} & {14} & {14} & {17} \\
\hline
{\textbf{Bypass}} & {\checkmark} & {\checkmark} & {\checkmark} & {\checkmark} & {\checkmark} & {-} & {-} & {-} & {-} & {-} \\
\hline
\end{tabular}

\label{tab:dnn_sk}
\end{table}

\begin{table}[h!]
\vspace{0.0em}
\caption{The 5 model variants of MobileNetV2 ~\cite{mobilenetv2} using different channel scaling factors and input resolutions.}
\vspace{-1.0em}
\def\arraystretch{1.2}
\centering
\scriptsize
\begin{tabular}{|c|c|c|c|c|c|}
\hline
{\textbf{DNN}} & {\textbf{V-Model 1}} & {\textbf{V-Model 2}} & {\textbf{V-Model 3}} & {\textbf{V-Model 4}} & {\textbf{V-Model 5}} \\
\hline
\multirow{1}{*}{Resolution} & {128} & {128} & {224} & {224} & {224}\\
\hline
\multirow{1}{*}{Channel scaling} & {0.5} & {1.0} & {0.5} & {1.0} & {1.4}\\
\hline
\end{tabular}
\label{tab:dnn_mb}
\vspace{-1em}
\end{table}

\textbf{Validation of the Predicted Energy Consumption.}
We compare the \textit{Chip Predictor}'s predicted energy with the measured ones from 3 edge devices, including Ultra96 FPGA~\cite{ultra96} (edge FPGA), edge TPU~\cite{edgetpu}, and Jetson TX2 (edge GPU)~\cite{edgegpu}) under the same settings (see Table~\ref{tab:val_setting}). 

\begin{figure}[!b]
\vspace{-1em}
    \centerline{\includegraphics[width=1.0\linewidth]{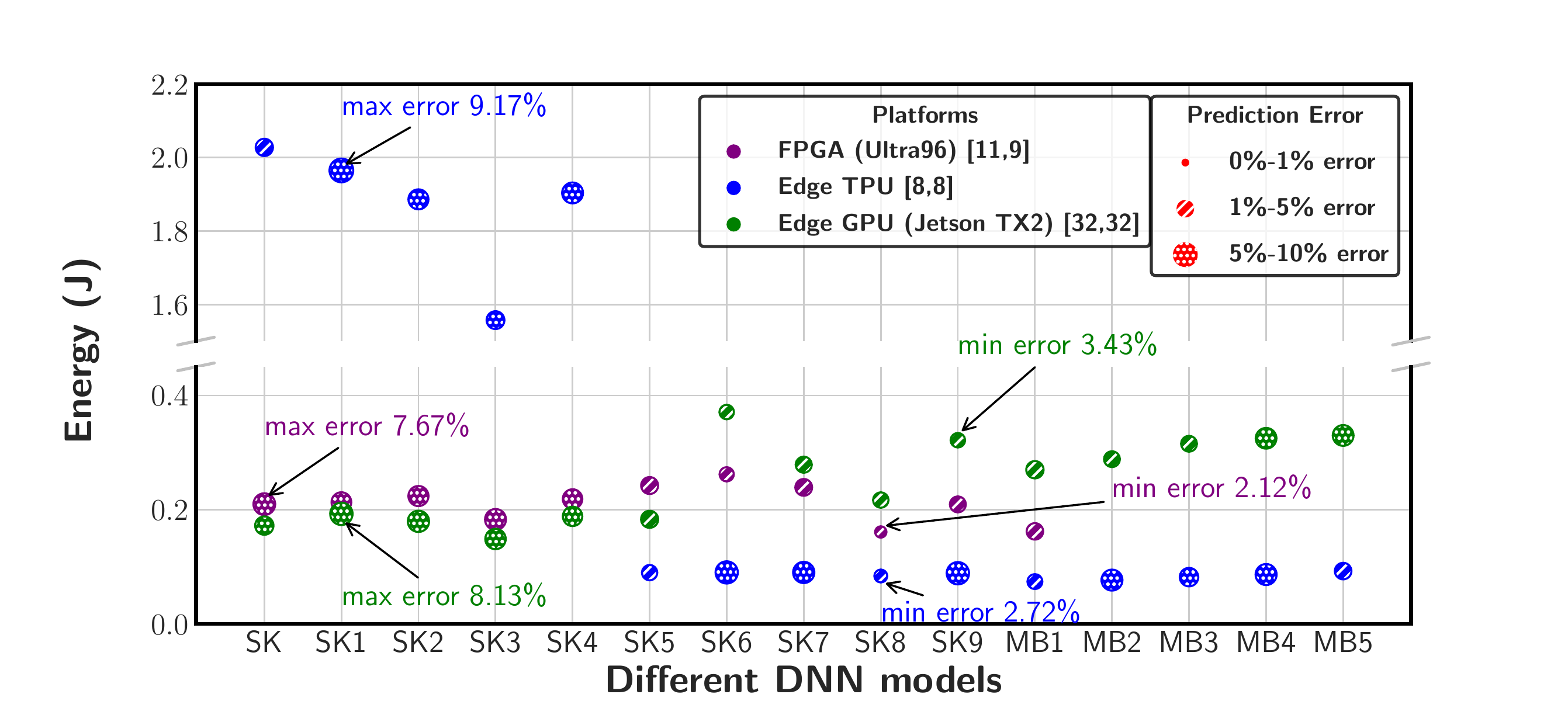}}
    \vspace{-1.2em}
    \caption{\textbf{The energy prediction error of the \textit{Chip Predictor} when using the 15 DNN models in Table~\ref{tab:dnn_sk} and Table~\ref{tab:dnn_mb} running on 3 edge devices including an edge FPGA ~\cite{ultra96}, Edge TPU~\cite{edgetpu}, and edge GPU~\cite{edgegpu}.}}
    \label{fig:val_energy}
\end{figure}

Fig. ~\ref{fig:val_energy} summarizes the validation results, and shows that \textbf{the maximum prediction error of our \textit{Chip Predictor} is $\boldsymbol{9.17\%}$ for all 15 DNN models across 3 platforms}. 
Specifically, the prediction error ranges from $0.89\%$ to $8.13\%$, $2.12\%$ to $7.67\%$, and $2.72\%$ to $9.17\%$, for the cases using the edge GPU, the edge FPGA, and the edge TPU, respectively, and the corresponding average prediction error is $5.40\%$, $5.20\%$, and $6.05\%$, respectively. 
We notice the energy consumption of the SkyNet and SK1-SK4 models are relatively large using the edge TPU. The reason is that these models contain unsupported operations (e.g., short-cut paths and feature map reorganization~\cite{zhang2019skynet}) that need to be handled by the embedded CPU instead of the optimized tensor unit with higher efficiency. 

\begin{figure}[!t]
    \vspace{-1.0em}
    \centering
    \subfloat[]{\includegraphics[width=40mm]{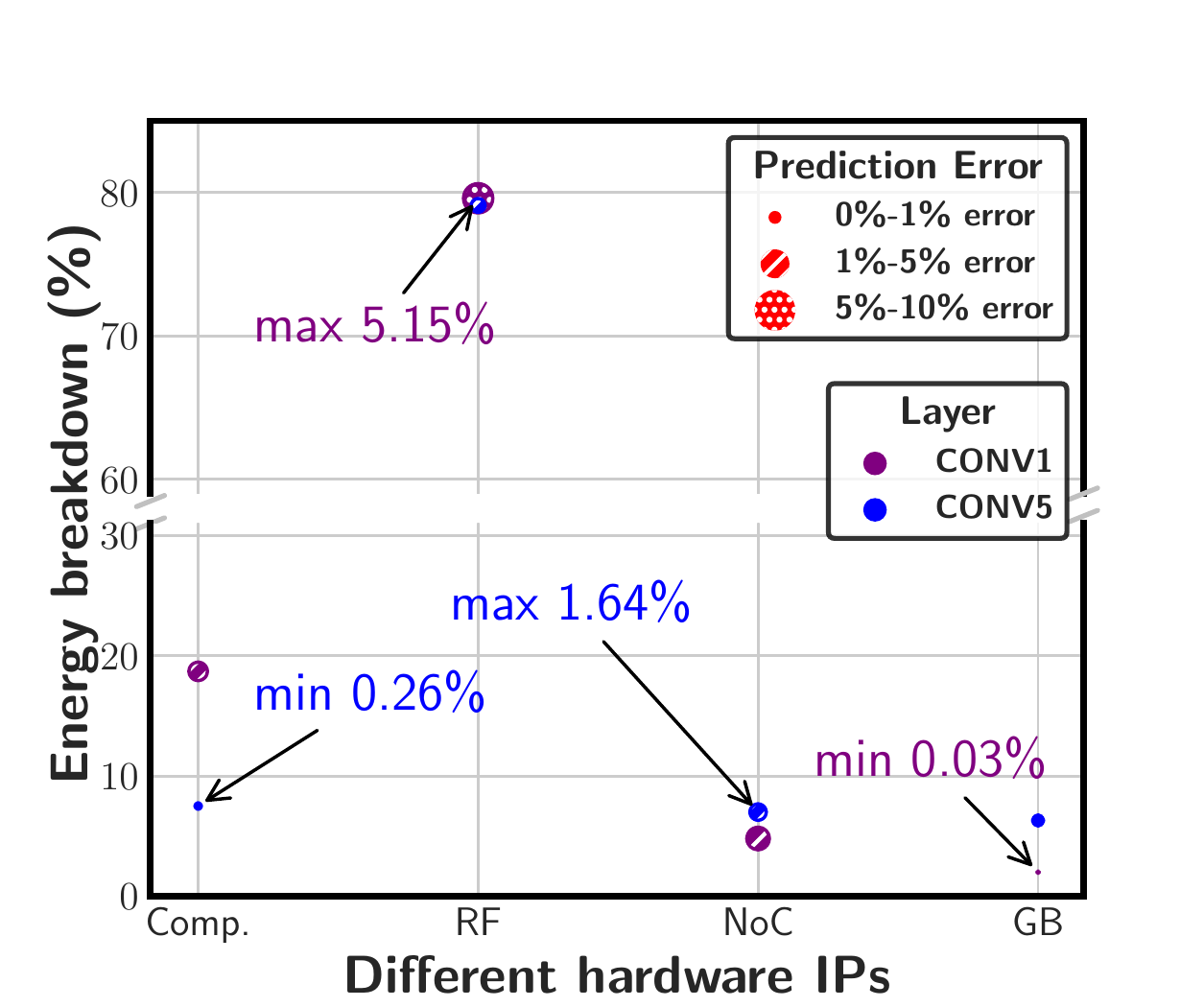} \label{subfig:val_energy_eyeriss_a}}
    \subfloat[]{\includegraphics[width=42mm]{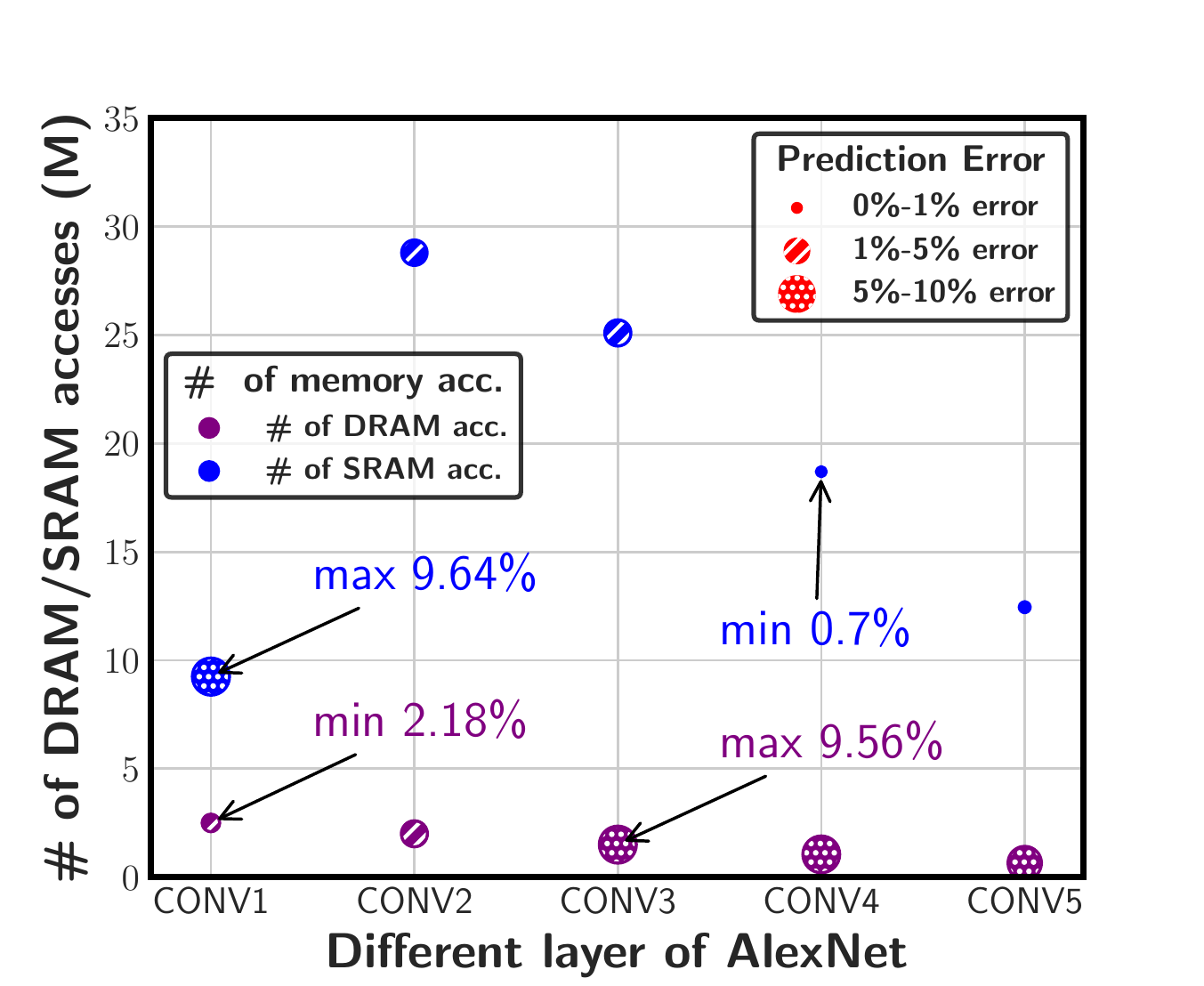} \label{subfig:val_energy_eyeriss_b}}
    \vspace{-1.2em}
    \caption{\textbf{
    The \textit{Chip Predictor}'s energy prediction error considering the Eyeriss  architecture~\cite{chen2017eyeriss}: (a) The energy breakdown for AlexNet's 1st and 5th convolutional layers and (b) the $\#$ of DRAM and SRAM accesses of convolutional layers.}}
    \vspace{-0.5em}
    \label{fig:val_energy_eyeriss}
\end{figure}

In Fig.~\ref{fig:val_energy_eyeriss} and Table~\ref{tab:val_energy_shidiannao}, we validate the proposed \textit{Chip Predictor} by comparing it to 2 state-of-the-art ASIC-based accelerators: Eyeriss~\cite{eyeriss} and ShiDianNao \cite{du2015shidiannao}.
\ul{For Eyeriss}~\cite{eyeriss}, we first compare the predicted energy breakdown of the first and fifth convolutional layers of AlexNet, of which the maximum error is $5.15\%$ and $1.64\%$, respectively, as shown in Fig.~\ref{fig:val_energy_eyeriss} (a). Since the memory accesses dominate the energy consumption~\cite{eyeriss}, we further compare the number of DRAM and SRAM accesses. In Fig.~\ref{fig:val_energy_eyeriss} (b), we present the error between the predicted and Eyeriss's paper-reported results. The relatively large error of SRAM accesses in the first convolutional layer is caused by the unsupported large stride number (4 in this case) since our predictor only  considers the commonly used strides of 1 and 2 for simplicity. 
Note our \textit{Predictor} can be straightforwardly extended to include other stride values. Note that the prediction errors of DRAM accesses in the last three layers are relatively large, because the input data are compressed to save DRAM accesses in ~\cite{eyeriss} and we lack their information regarding the input data sparsity. The validation results over \ul{ShiDianNao}~\cite{du2015shidiannao} are listed in Table~\ref{tab:val_energy_shidiannao}. 
By showing the average energy over 10 DNN benchmarks of the 4 IPs in~\cite{du2015shidiannao}, we verify the maximum prediction error is  9.59\%, where the error is mainly due to the difference between our adopted commercial CMOS IP library and the one used in ~\cite{du2015shidiannao}. 

\begin{table}[h!]
\vspace{-0.0em}
\caption{\textbf{The energy prediction error of the \textit{Chip Predictor} when using the architecture of ShiDianNao~\cite{du2015shidiannao}: The energy breakdown over 10 benchmarks.}}
\vspace{-1.0em}
\def\arraystretch{1.2}
\centering
\scriptsize
\begin{tabular}{|c|c|c|c|c|}
\hline
{\textbf{IP}} & {\textbf{Computation}} & {\textbf{Input SRAM}} & {\textbf{Output SRAM}} & {\textbf{Weight SRAM}}  \\
\hline
\multirow{1}{*}{Predicted (\%)} & {89.2} & {7.4} & {1.7} & {1.6}\\
\hline
\multirow{1}{*}{Paper-reported (\%)} & {89.0} & {8.0} & {1.6} & {1.5} \\
\hline
\multirow{1}{*}{\textbf{Prediction error}} & {\textbf{0.35\%}} & {\textbf{-7.19\%}} & \textbf{{9.59\%}} & {\textbf{7.87\%}} \\
\hline
\end{tabular}
\vspace{-0.7em}
\label{tab:val_energy_shidiannao}
\end{table}

\begin{figure}[h]
\vspace{-0.5 em}
    \centerline{\includegraphics[width=1.0\linewidth]{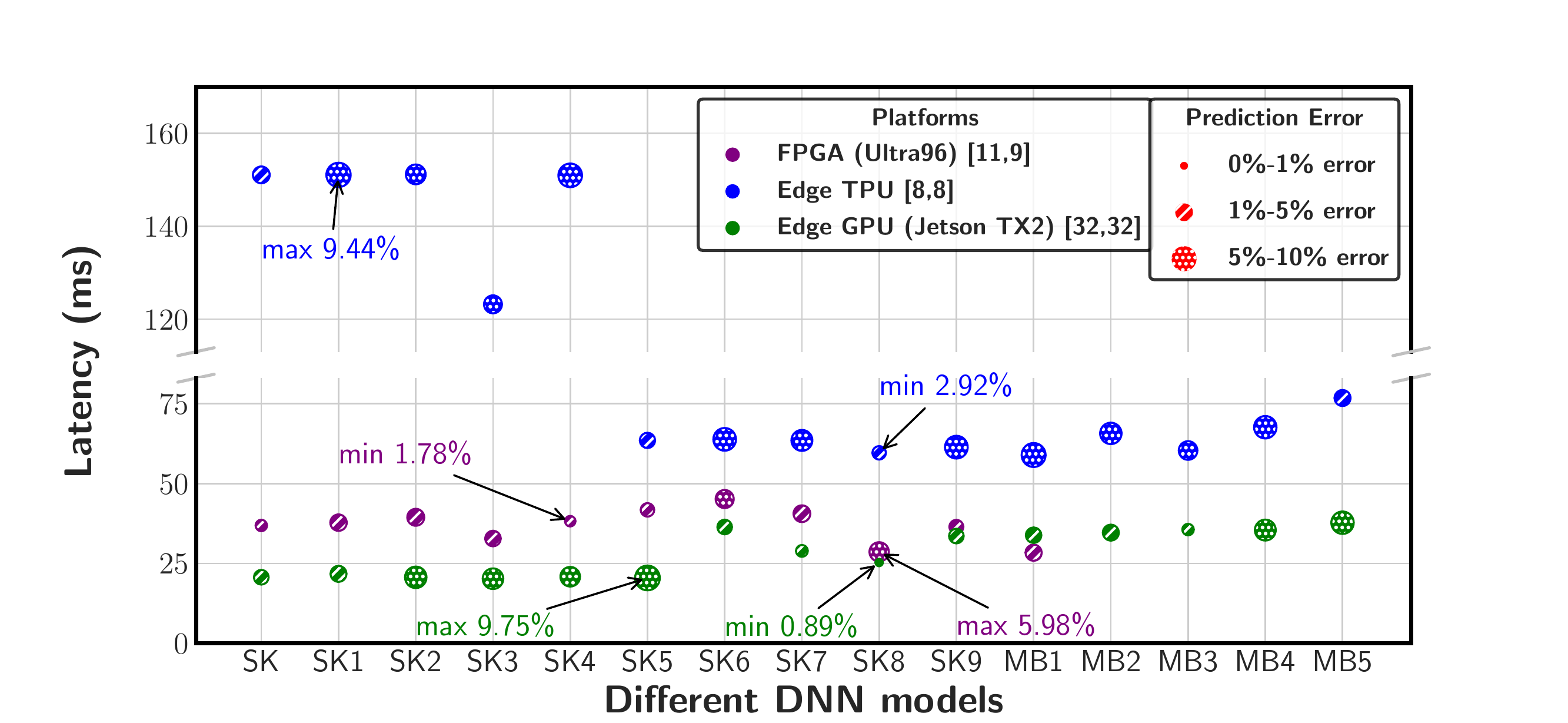}}
    \vspace{-1.2em}
    \caption{\textbf{The latency prediction error of the \textit{Chip Predictor} when using 15 DNN models on 3 edge devices including Ultra96 FPGA board, Edge TPU, and Jetson TX2 (edge GPU).}}
     \vspace{-0.5em}
    \label{fig:val_lat}
\end{figure}

\textbf{Validation of the Predicted Latency.} 
The latency prediction of the \textit{Chip Predictor} is validated over the measured results of the same 15 DNN models and 3 edge devices and shown in Fig.~\ref{fig:val_lat}. he edge GPU, the Ultra96 FPGA board, and the edge TPU, respectively, and the corresponding average prediction error is $4.85\%$, $3.73\%$, and $6.57\%$, respectively. 
\textbf{The maximum latency prediction error of our \textit{Chip Predictor} is $\boldsymbol{9.75\%}$}. Specifically, the prediction errors range from $0.89\%$ to $9.75\%$, $1.78\%$ to $5.98\%$, and $2.92\%$ to $9.44\%$, when using the edge GPU, the edge FPGA, and the edge TPU, respectively. The corresponding average prediction error is $4.85\%$, $3.73\%$, and $6.57\%$, respectively. 
Similar to the case in Fig.~\ref{fig:val_energy}, the latency of the SkyNet and SK1-SK4 models when using the edge TPU are relatively large because of the unsupported operations. 

Table~\ref{tab:val_eyeriss_lat} summarizes the latency prediction when running the 5 convolutional layers of AlexNet on \ul{Eyeriss}~\cite{eyeriss} with the largest error peaking at $4.12\%$. The predicted latency generated by the proposed \textit{Chip Predictor} is smaller than the paper-reported results, as \textit{Chip Predictor} does not consider the special scenario when the accelerator needs to access memory multiple times for one single wordline of data. Such a scenario only happens when one wordline of data is physically stored in multiple wordlines of the memory. The \textit{Chip Predictor} can be extended to include such a case by configuring corresponding memory data arrangements.


\begin{table}[h!]
\caption{\textbf{The latency prediction error of the \textit{Chip Predictor} when using the architecture of Eyeriss~\cite{eyeriss}: The latency when processing the 5 convolutional layers of AlexNet}.}
\vspace{-1.0em}
\def\arraystretch{1.2}
\centering
\scriptsize
\begin{tabular}{|c|c|c|c|c|c|}
\hline
{\textbf{AlexNet Layer}} & {\textbf{CONV1}} & {\textbf{CONV2}} & {\textbf{CONV3}} & {\textbf{CONV4}} & {\textbf{CONV5}} \\
\hline
\multirow{1}{*}{Predicted latency (ms)} & {16.04} & {37.58} & {21.09} & {15.59} & {9.79}\\
\hline
\multirow{1}{*}{Paper-reported latency (ms)} & {16.5} & {39.2} & {21.8} & {16} & {10}\\
\hline
\multirow{1}{*}{\textbf{Prediction error}} & \textbf{{-2.88\%}} & {\textbf{-4.12\%}} & \textbf{{-3.24\%}} & \textbf{{-2.56\%}} & \textbf{{-2.14\%}}\\
\hline
\end{tabular}
\vspace{-0.5em}
\label{tab:val_eyeriss_lat}
\end{table}






\textbf{Validation of the Predicted Resource Consumption.} 
Table~\ref{tab:val_resource} summarizes the \textit{Chip Predictor}'s predicted resource consumption based on the experiments using \ul{Ultra96 FPGA}~\cite{ultra96}. Specifically, the predicted resource consumption for the 2 critical on-chip resources of FPGAs, DSP48E and BRAM18K, is validated against those obtained from the post-implementation utilization reports, and has a corresponding prediction error of smaller than $4.2\%$ and $3.2\%$, respectively. Note that the DSP48Es are the embedded multipliers in the FPGA and the BRAM18K is the main on-chip memory resource in the FPGA. In addition, the 6 cases, i.e., Bg. 1-6 in Table~\ref{tab:val_resource}, correspond to 6 designs under varied resource budgets. 
For the validation over ASIC-based DNN accelerators, we consider the MAC utilization as the validation metric. Estimated MAC utilization among the 5 convolutional layers of AlexNet is the same as the paper-reported ones in \ul{Eyeriss}~\cite{eyeriss}, because the MAC utilization is only determined by the parallelism level of the PE array in Eyeriss.


\begin{table}[h!]
\vspace{-0.5em}
\caption{\textbf{
The resource consumption prediction error of the \textit{Chip Predictor}'s on the Ultra96 FPGA board when considering 6 different designs given 6 different resource budgets.}}
\vspace{-1.0em}
\def\arraystretch{1.2}
\centering
\scriptsize
\begin{tabular}{|c|c|c|c|c|c|c|c|}
\hline
{\textbf{Resource type}}& {Val.}  & {\textbf{Bg. 1}} & {\textbf{Bg. 2}} & {\textbf{Bg. 3}} & {\textbf{Bg. 4}} & {\textbf{Bg. 5}} & {\textbf{Bg. 6}} \\
\hline
\multirow{3}{*}{DSP48E} & Predicted & 35 & 69 & 141 & 213 & 285 & 331 \\
 & Measured & 36 & 72 & 144 & 216 & 288 & 360 \\
 & \textbf{Error} & \textbf{-3.2\%} & \textbf{-4.2\%} & \textbf{-2.4\%} & \textbf{-1.2\%} & \textbf{-1.0\%} & \textbf{-0.8\%} \\
\hline
\multirow{3}{*}{BRAM18K} & Predicted & 65 & 87 & 175 & 265 & 354 & 446\\
 & Measured & 64 & 86 & 173 & 259 & 346 & 432 \\
 & \textbf{Error} & \textbf{+1.0\%} & \textbf{+0.8\%} & \textbf{+1.2\%} & \textbf{+2.2\%} & \textbf{+2.4\%} & \textbf{+3.2\%} \\
\hline
\end{tabular}
\vspace{-2.7em}
\label{tab:val_resource}
\end{table}


\vspace{0.5em}
\subsection{Evaluation for the \textit{Chip Builder} and \textit{AutoDNNchip}}
\label{subsec:dse}
In this subsection, we evaluate the performance of the proposed \textit{Chip Builder}, which makes use of the time-efficient and accurate \textit{Chip Predictor} to perform an effective two-stage DSE efficiently, and \textit{AutoDNNchip}. Specifically, we study the performance of the resulting DNN accelerators generated and optimized by the \textit{Chip Builder} and \textit{AutoDNNchip}. \ul{First}, we show experiment results to visualize the \textit{Chip Builder}'s two-stage DSE process; \ul{Second}, we study the performance improvement resulted from the \textit{Chip Builder}'s second-stage IP-pipeline co-optimization in terms of bottleneck blocks' latency and idle cycle reduction; \ul{Finally}, we validate the effectiveness of the \textit{AutoDNNchip} by comparing the performance of its generated FPGA- and ASIC-based accelerators (i.e., the corre-

\noindent sponding RTL implementation) with that of state-of-the-art designs under the same conditions. 



\vspace{0.2em}
\textbf{Evaluation Setup.} In this set of experiments, we consider the application-driven specifications and constraints summarized in Table ~\ref{tab:dse_setting}, where the throughput requirement and power budget are set to meet real-time applications of visual recognition (e.g., image classification and object detection~\cite{ren2015faster}) on edge devices. For the FPGA-based accelerator design, we use a state-of-the-art edge device Ultra96 FPGA board ~\cite{ultra96} whose resource budget is fixed; for the ASIC-based accelerator design, we evaluate our generated designs through RTL simulations. Regarding the design space exploration, we use \textit{Algorithm} ~\ref{alg:sim} to perform the accelerator design optimization, considering the architecture/dataflow search space and the design factors in Table~\ref{tab:design_space} for the IP/dataflow design.

\begin{figure}[h]
\vspace{-0em}
    \centerline{\includegraphics[width=1.0\linewidth]{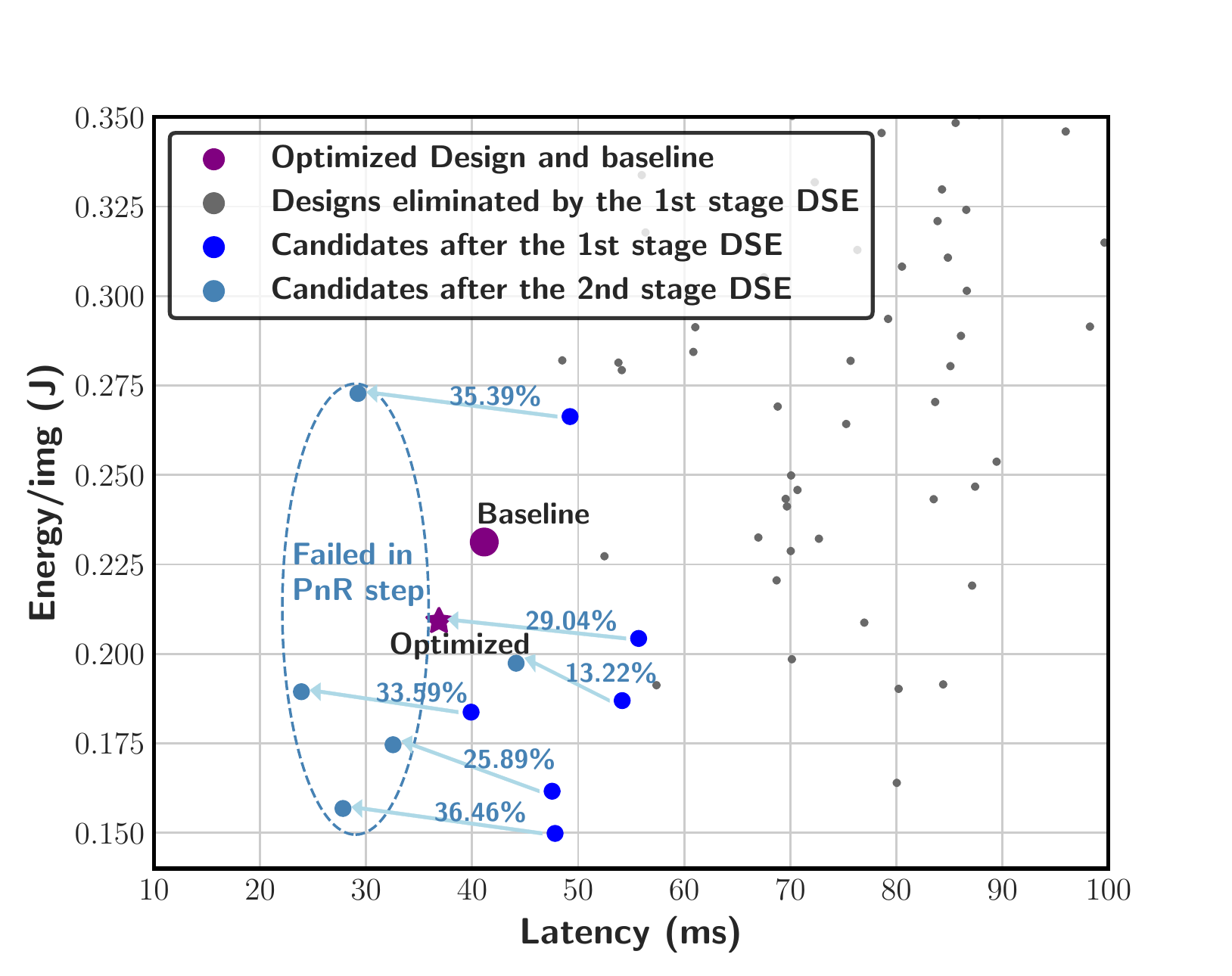}}
    \vspace{-1.0em}
    \caption{\textbf{Visualizing the energy consumption per image and processing latency of the resulting designs from the \textit{Chip Builder}'s 1st and 2nd stage optimization, when  using \textit{AutoDNNchip} to design an FPGA-based accelerator for meeting the  performance of a state-of-the-art design ~\cite{zhang2019skynet} given the same performance specification/constraint, FPGA board, DNN model, and dataset (see Table ~\ref{tab:dse_setting}).
    }}
    \vspace{-0.5em}
    \label{fig:dse_fpga}
\end{figure}

\vspace{0.5em}
\textbf{Visualizing the \textit{Chip Builder}'s Two-stage DSE Process.} For demonstrating the effectiveness of the \textit{Chip Builder}'s two-stage DSE engine, 
here we visualize 
the DSE process, when using \textit{AutoDNNchip} to design an FPGA-based accelerator for achieving competitive performance as the award winning state-of-the-art design in ~\cite{zhang2019skynet} given the same target performance specification/constraint, FPGA board, DNN model, and dataset. The FPGA measured energy consumption of both the resulting design from the \textit{AutoDNNchip} and the reported one of ~\cite{zhang2019skynet} are marked in purple in Fig.~\ref{fig:dse_fpga}. It demonstrates that: (1) the DSE engine of the \textit{Chip Builder} can effectively trim down the design choices and generate optimized designs with better performance compared to the state-of-the-art design published in \cite{zhang2019skynet}.
\textbf{Without humans in the loop, the \textit{AutoDNNchip} can indeed automatically generate DNN accelerators that achieve optimized performance}; (2) most of the design choices can be efficiently ruled out by \ul{the 1st stage} of the DSE engine, i.e., the early stage exploration based on the \textit{Chip Predictor}'s coarse-grained analytical performance estimation; and (3) \ul{the 2nd stage} IP-pipeline co-optimization of the \textit{Chip Builder} can effectively boost (up to 36.46\% improvement and an average of 28.92\% improvement) the performance, i.e., throughput of the DNN accelerators here, as compared to that of the  designs resulted from the 1st stage DSE. The final generated design candidates in the HLS code format will be passed to Vivado~\cite{vivado_HLS} for implementation. Then, we eliminate the designs that fail in the PnR step as shown in Fig.~\ref{fig:dse_fpga} and find an optimal design from the remaining ones. As a reference point, the 1st stage DSE takes about 0.65 ms for each design point and only 0.8 hour for exploring a total of 4.6 million design points when running on an Intel Core i5 CPU with a single thread, thanks to the analytical nature of the \textit{Chip Predictor}.

\begin{figure}[!b]
    \vspace{-0.5em}
    \centerline{\includegraphics[width=1.05\linewidth]{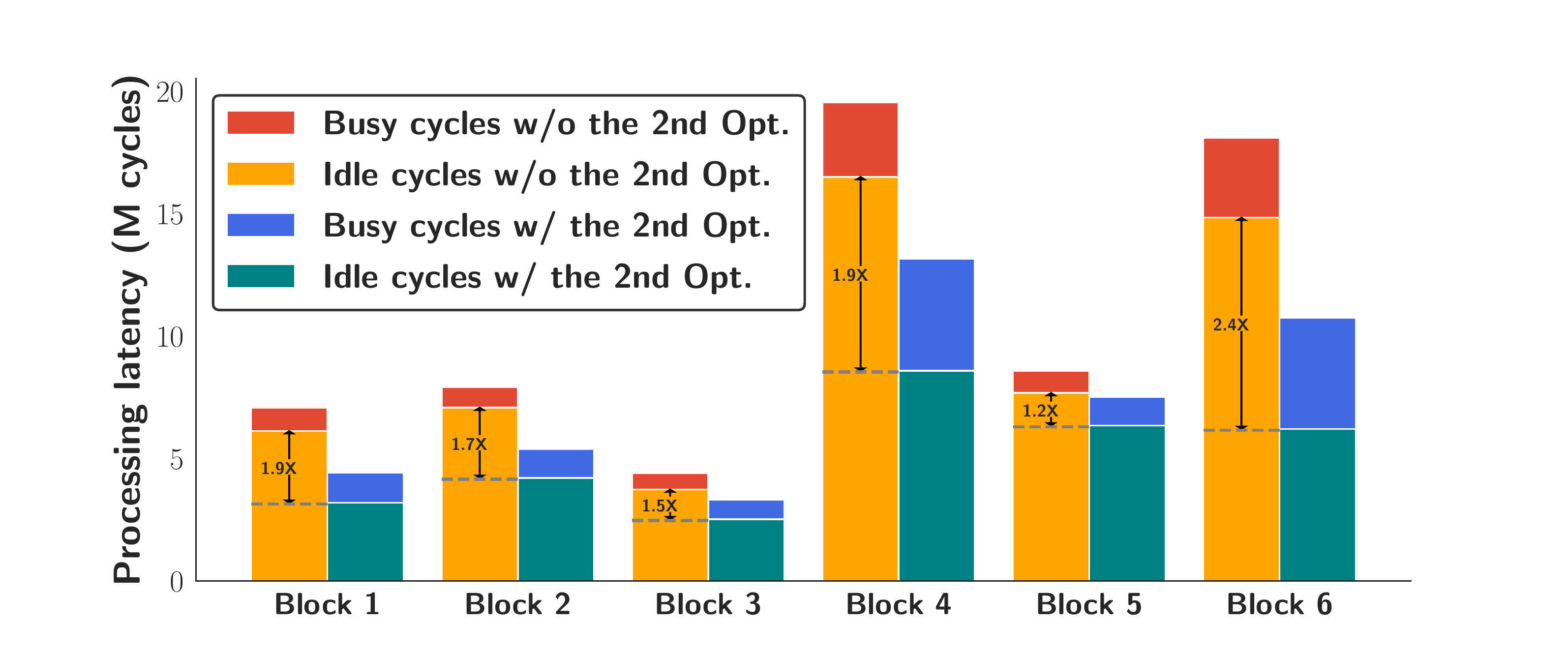}}
    \vspace{-1.0em}
    \caption{\textbf{The busy and idle cycles of the bottleneck IPs in SkyNet's 6 different blocks, before and after conducting the \textit{Chip Builder}'s 2nd-stage IP-pipeline co-optimization, when using the \textit{AutoDNNchip} to generate designs for the Ultra96 FPGA board with the same target performance as \cite{zhang2019skynet}.}}
    \label{fig:idle_cycles_fpga}
\end{figure}

\begin{table}[h!]
\vspace{-0.5em}
\caption{The considered application-driven specifications (i.e., throughput requirement) and constraints (i.e., power and resource budget) when evaluating the \textit{Chip Builder}'s generated FPGA- and ASIC-based DNN accelerators.}
\vspace{-1.0em}
\def\arraystretch{1.2}
\centering
\scriptsize
\begin{tabular}{|c|c|c|c|c|}
\hline
{\textbf{Target Back-end}} & {\textbf{Application}} & {\textbf{Opt. Obj.}} & {\textbf{Th./P. Req.}}  & {\textbf{Res. Budget}}\\
\hline
\multirow{2}{*}{\textbf{Ultra96 FPGA}} & \multirow{2}{*}{Object Detection} & \multirow{2}{*}{E, L} & {20FPS}  & {DSP=360, FF=141120}\\ 
 & &  & {10W} & {LUT=70560, BRAM=432}\\ 
\hline
\multirow{2}{*}{\textbf{ASIC}} & \multirow{2}{*}{Vision Tasks \cite{du2015shidiannao}} & \multirow{2}{*}{E, L} & {15FPS}  & {On-chip SRAM=128KB}\\
 &  &  & 600mW  & {\# of MAC units=64}\\
 \hline

\end{tabular}

 \label{tab:dse_setting}
\end{table}
\vspace{-1.0em}
\textbf{Evaluation of the \textit{Chip Builder}'s 2nd-stage Optimization.} Fig.~\ref{fig:idle_cycles_fpga} summarizes the evaluation experiments for the \textit{Chip Builder}'s 2nd-stage Optimization process. As described in Section \ref{sec:dse}, this stage targets an IP-pipeline co-optimization and thus can lead to more balanced pipeline and more efficient resource allocation. From Fig.~\ref{fig:idle_cycles_fpga}, we can see that the \textit{Chip Builder}'s 2nd-stage optimization can achieve up to 2.4$\times$ idle cycles reduction, when optimizing the design of SkyNet's 6 blocks~\cite{zhang2019skynet} on the Ultra96 edge FPGA board ~\cite{ultra96}.





\begin{figure}[h]
\vspace{-0.5em}
    \centerline{\includegraphics[width=1\linewidth]{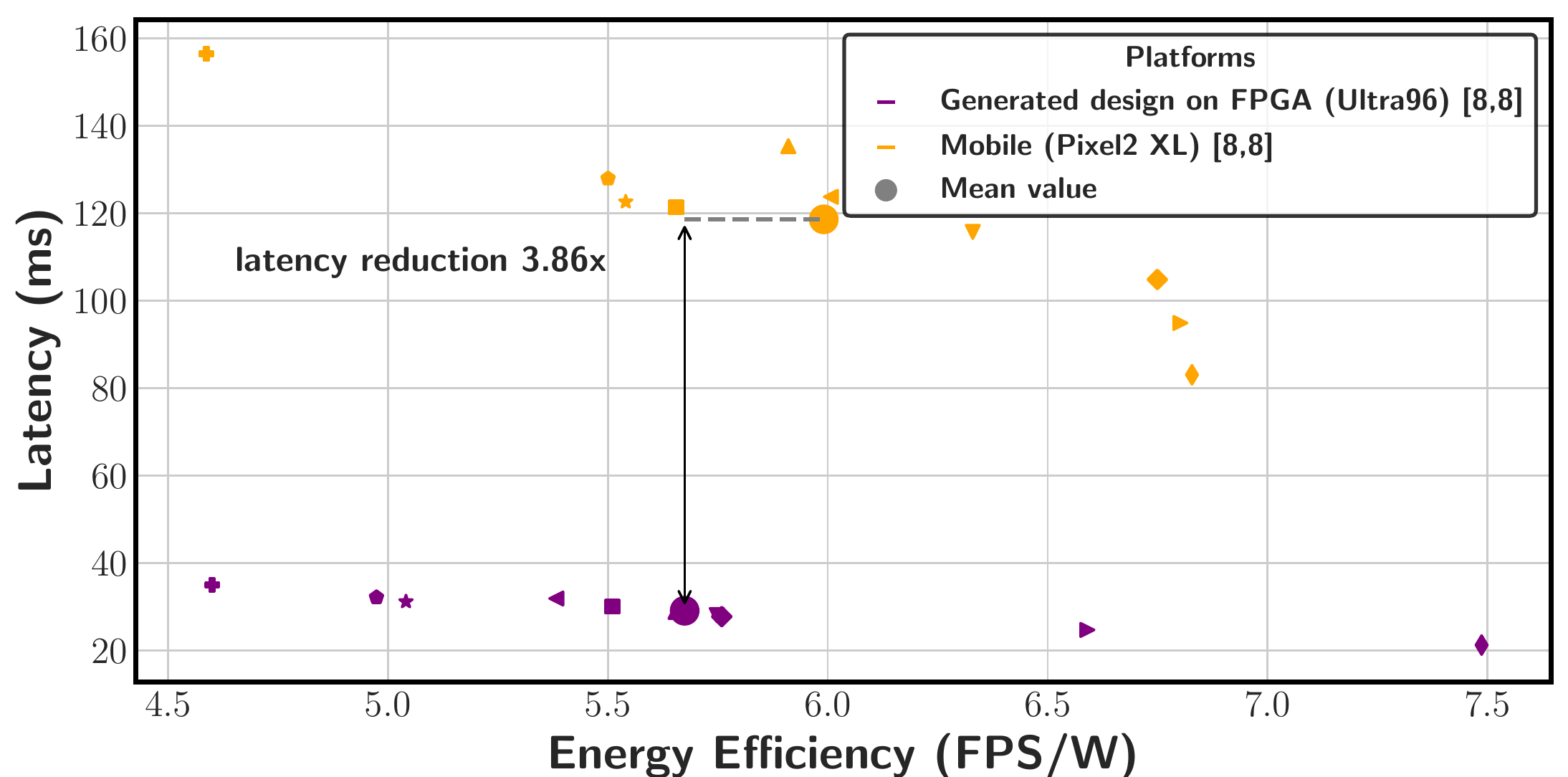}}
    \vspace{-0.8em}
    \caption{\textbf{Processing latency and energy efficiency on Ultra96 FPGA compared with a mobile device (Pixel2 XL ) on 10 compact DNN models}}
    \vspace{-2em}
    \label{fig:pareto}
\end{figure}



\textbf{Evaluation of \textit{AutoDNNchip}'s Generated FPGA-based Accelerators.} Fig. \ref{fig:dse_fpga} shows that the DNN accelerator which is generated by \textit{AutoDNNchip} can apparently outperform the recent award-winning design \cite{zhang2019skynet}. We further conduct another set of experiments to compare the performance of \textit{AutoDNNchip}'s generated DNN accelerators on the Ultra96 FPGA board with that of a mobile CPU (Pixel2 XL~\cite{mobile}), when both designs (1) adopt the settings in Table~\ref{tab:val_setting}, using the same bit precision and the 10 DNN models in Table~\ref{tab:dnn_sk}, and (2) try to minimize the latency for considering time-critical applications. Note that the DNN mapping to the mobile CPU is optimized using Tensorflow Lite~\cite{tflite}. Fig. \ref{fig:pareto} illustrates the corresponding latency vs. energy efficiency, where the results under the same DNN models are marked with makers of the same shape. We can see that \textit{AutoDNNchip} generated accelerators consistently achieve smaller latency than the baselines under the same DNN model and settings while having similar (<15\% difference) energy efficiency. Specifically, \textit{AutoDNNchip} generated accelerators achieve an average latency reduction of 3.86$\times$ while having slightly better (10\%) or worse (differs <15\%) energy efficiency, demonstrating the effectiveness of \textit{AutoDNNchip} in generating optimized FPGA-based accelerators.

\begin{figure}[!t]
    \centerline{\includegraphics[width=0.95\linewidth]{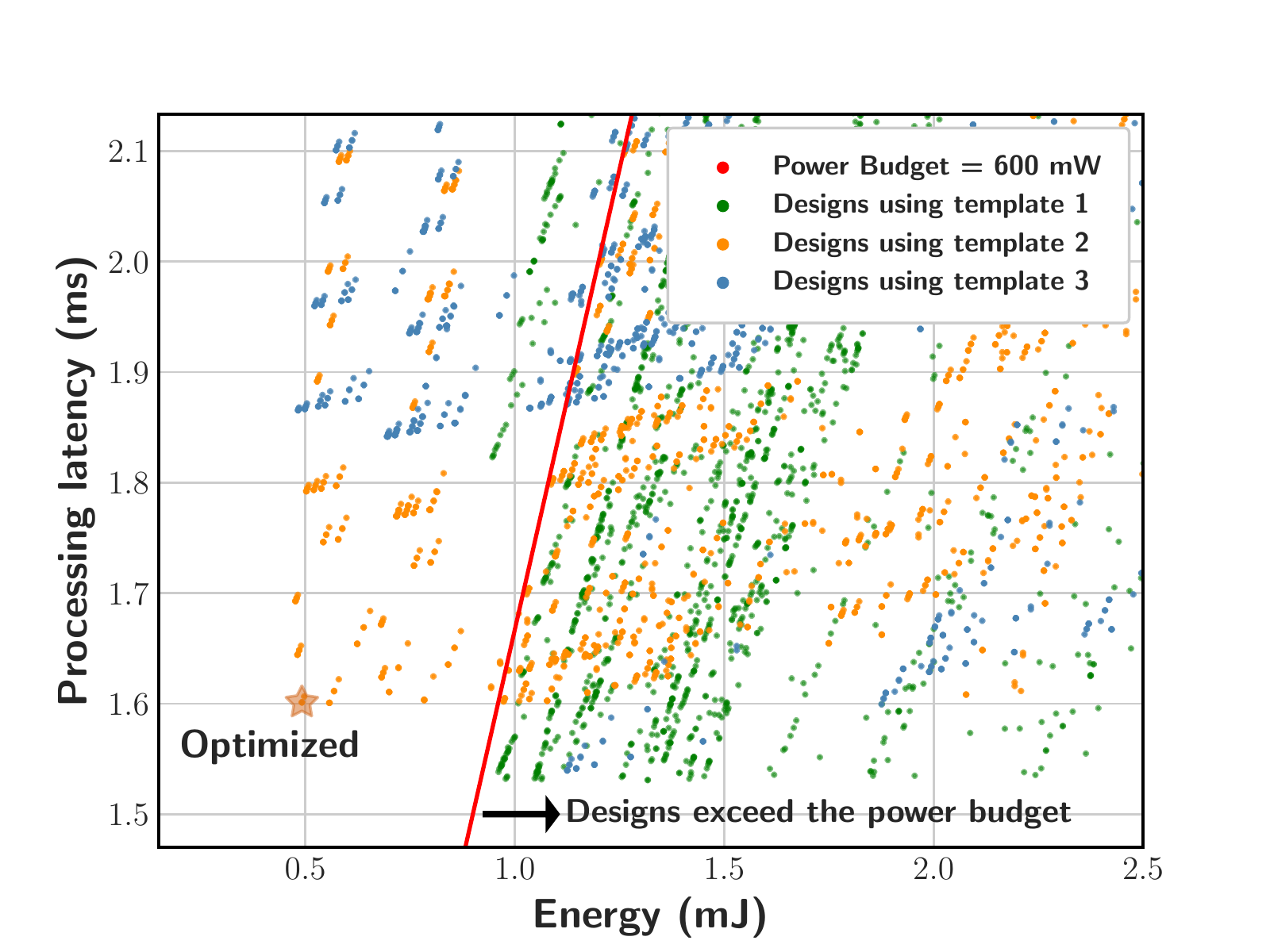}}
    \vspace{-0.8em}
    \caption{\textbf{Visualizing the latency vs. energy consumption per image of the ASIC-based accelerators in the  design space pool, when using \textit{AutoDNNchip} to design an ASIC-based accelerator for meeting the performance of a state-of-the-art ASIC-based accelerator~\cite{du2015shidiannao}, with both having the same performance constraints, DNN model, and dataset (see Table ~\ref{tab:dse_setting}).
    }}
    \label{fig:dse_asic}
\end{figure}

\begin{figure}[h]
\vspace{-0.5em}
    \centerline{\includegraphics[width=1\linewidth]{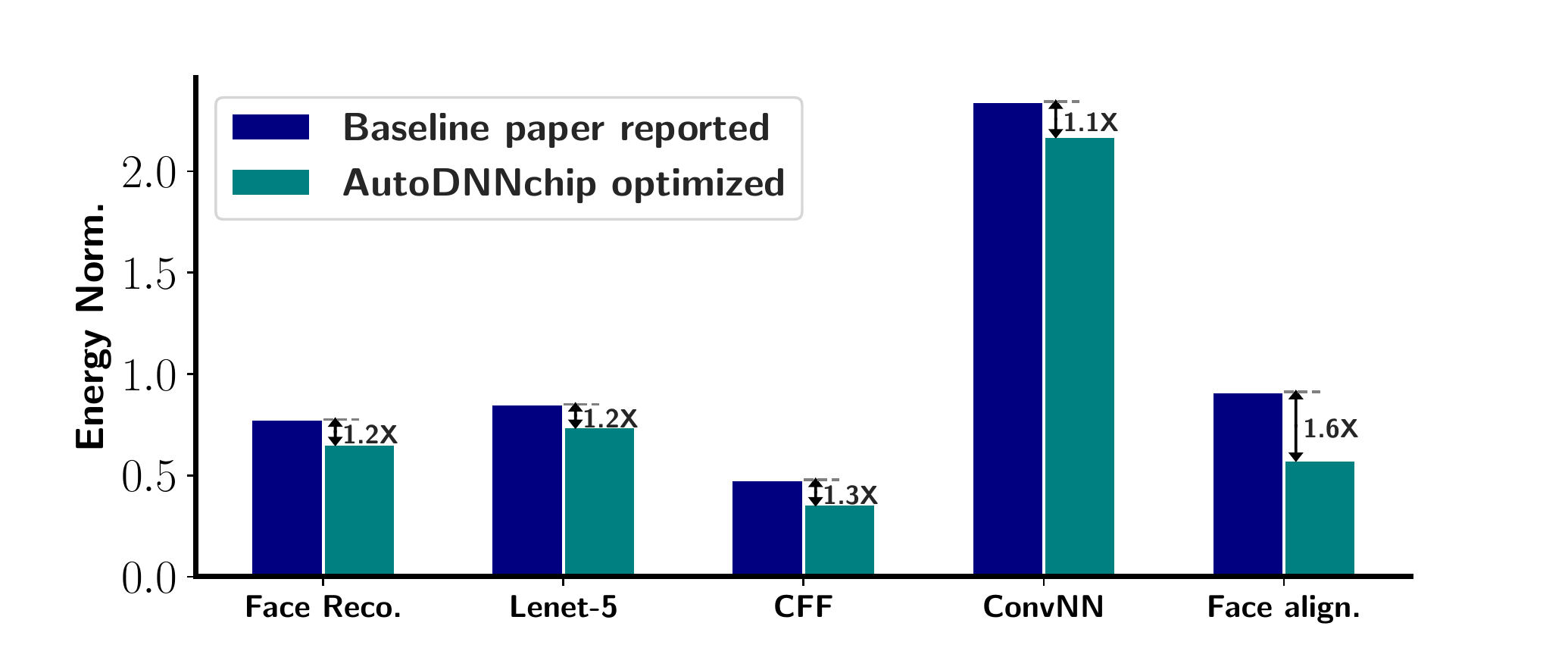}}
    \vspace{-0.8em}
    \caption{\textbf{Comparing the normalized energy consumption between the \textit{AutoDNNchip} generated ASIC-based accelerators and \cite{du2015shidiannao}, when accelerating 5 shallow neural networks under the same throughput requirement. }}
    \vspace{-0.5em}
    \label{fig:asic_comp}
\end{figure}

\textbf{Evaluation of \textit{AutoDNNchip}'s Generated ASIC-based Accelerators.} Fig.\ref{fig:dse_asic} illustrates that \textit{AutoDNNchip} indeed can generate ASIC-based accelerator that leads to an optimal tradeoff between latency and energy consumption by visualizing the latency vs. energy consumption of the generated accelerators, where dots with different colors correspond to designs based on different hardware templates. Furthermore, we evaluate the performance of \textit{AutoDNNchip} generated ASIC-based accelerators by comparing their energy consumption with that of a state-of-the-art ASIC-based accelerator \cite{du2015shidiannao} based on 5 shallow neural networks, which are used in \cite{du2015shidiannao} for performance evaluation, given both having the same throughput constraint as shown in Table ~\ref{tab:dse_setting}. Fig. \ref{fig:asic_comp} shows the comparison, where all the energy consumption in Fig.\ref{fig:asic_comp} are obtained from RTL implementation and simulation. We can see that \textit{AutoDNNchip} generated ASIC-based accelerators consistently outperform \cite{du2015shidiannao} in all the 5 networks with energy consumption improvement ranging from 7.9\% to 58.3\%, demonstrating the effectiveness of \textit{AutoDNNchip} in generating optimized ASIC-based accelerators.

For the aforementioned set of experiments, We first use the application-driven performance and constraints (see Table~\ref{tab:dse_setting}) to perform design space exploration and then validate the generated designs using RTL simulations adopting the same clock frequency (1 GHz) and technology (65nm) as our baseline~\cite{du2015shidiannao}. Specifically, the DSE process optimizes the accelerators' energy-delay-product, and considering different: (1) hardware templates with three different architectures \cite{TPU, du2015shidiannao, eyeriss} (denoted as template 1/2/3 in Fig.~\ref{fig:dse_asic}), (2) memory size and \# of PEs within the resource constraint, (3) mem-

\noindent ory allocation (i.e., input/weight/output buffer), and (4) memory accesses and reuse patterns.

\section{Conclusions}
To close the gap between the growing demand for DNN accelerators with various specifications and the time consuming and challenging DNN accelerator design, we develop \textit{AutoDNNchip} which can automatically generate both FPGA- and ASIC-based DNN accelerators. Experiments using over 20 DNN models and 4 platforms show that DNN accelerators generated by \textit{AutoDNNchip} outperform state-of-the-art designs by up to 3.86$\times$. Specifically, \textit{AutoDNNchip} is made possible by the proposed \textit{one-for-all design space description}, \textit{Chip Predictor}, and \textit{Chip Builder}. Experiments based on 15 DNN models and 4 platforms demonstrate that the \textit{Chip Predictor}'s prediction error is smaller than 10\% compared with real-measured ones, and the \textit{Chip Builder} can effectively and efficiently perform design space exploration and optimization. 

\vspace{-0.5em}
\begin{acks}
This work is supported in part by the NSF RTML grant 1937592 and NSF 1801865, the IBM-Illinois Center for Cognitive Computing System Research (C3SR), and XMotors.ai.
\end{acks}




\bibliographystyle{ieeetr}
\bibliography{ref}
\end{document}